\documentclass{article}
\usepackage{amsfonts,amssymb, amsmath}
\usepackage{float}
\usepackage{graphicx}
\usepackage{hyperref}

\usepackage[english]{babel}

\def\bq{ \begin{equation}}
\def\eq{ \end{equation}}

\def\a{{\alpha}}

\begin{document}


\title{ On inhomogeneous nonholonomic Bilimovich system}
\author{A.V. Borisov$^{(1)}$, E.A. Mikishanina$^{(2)}$, A.V. Tsiganov$^{(1)}$\\
$^{(1)}$ Steklov Mathematical Institute of Russian\\  Academy of Sciences, Moscow, Russia\\
 $^{(2)}$ Chuvash State University, Cheboksary, Russia\\
}%
\date{}
\maketitle

\begin{abstract}
We consider a simple motivating example of a non-Hamiltonian dynamical system with time-dependent constraints obtained by imposing rheonomic non-integrable Bilimovich's constraint on a freely rotating rigid body. Dynamics of this low-dimensional nonlinear nonautonomous dynamic system involves different kinds of stable and unstable attractors, quasi and strange attractors, compact and noncompact invariant attractive curves, etc. To study this cautionary example we apply the Poincar\'{e} map method to disambiguate and discover multiscale temporal dynamics, specifically the coarse-grained dynamics resulting from fast-scale nonlinear control via nonholonomic Bilimovich's constraint.
\end{abstract}

\section{Introduction}
Consider the motion of a rigid body under its inertia subjected to the rheonomic constraint
\bq\label{bil-c}
f=p-g(t)\,q-\a=0
\eq
where $\omega=(p,q,r)$ is an angular velocity vector in the body frame, $f(t)$ is an arbitrary function and $a$ is an arbitrary constant. Bilimovich studied this system with homogeneous constraint (\ref{bil-c}) at $\a = 0$ as a rheonomic generalization of the scleronomic Suslov system \cite{sus}.

In \cite{bil} Bilimovich suggested a mechanism with a rotating rod to physically implement constraint (\ref{bil-c}) at $\a=0$. In our opinion, this mechanism is unrealistic in contrary to the implementation of the Suslov constraint involving wheels with sharp edges rolling over a fixed sphere \cite{vag}. Other possible mechanical implementations of the similar nonintegrable constraints for angular velocity vector components are discussed in \cite{grig}.

There are many papers where the Suslov problem is studied by using various mathematical techniques, see \cite{drag,fer,fed09, nar14,mac,mah12,put} and references within. Therefore, it is reasonable to study its rheonomic generalization proposed by Bilimovich, despite the lack of physical implementation of the inhomogeneous constraint (\ref{bil-c}). According to \cite{bl, grab09, dl, kar, kol04,koz,nf72,sus} many nonholonomic systems have no physical implementations of constraints, but studies of abstract mathematical nonholonomic equations of motion provide for a better comprehension of locomotion generation, controllability, motion planning, trajectory tracking, etc.

According to \cite{bil} constraint (\ref{bil-c}) at $\a=0$ is the so-called ideal or perfect constraint, so herewith we have a dynamical system in three-dimensional phase space with one constraint (\ref{bil-c}) and one integral of motion
\begin{equation}\label{ham-bil}
T=\frac12(\mathbb I\omega,\omega)\,,
\end{equation}
where $\mathbb I$ is the inertia tensor of the body. Thus, at $\alpha=0$ we have a system integrable by quadratures, see \cite{bil,bt20} and references within.

 In this note we suppose that $\alpha\neq 0$, and $g(t)$ is a periodic function which allows us to study the inhomogeneous Bilimovich system by using the Poincar\'{e} map. This is an integral aspect of our understanding and analysis of nonlinear dynamical systems.

In \cite{poi} Henri Poincar\'{e} laid the foundations of what would become modern dynamical systems. In particular, he proposed to study the three-body problem by using the first recurrence map, or Poincar\'{e} map, characterizing the intersection of a periodic orbit in the state space of a continuous dynamical system with a lower-dimensional subspace transverse to the flow. Description of
 other qualitative methods for identifying dynamic behavior of a nonlinear system are given in the textbooks \cite{kuz,sh10,st15}.
Except for cases with the most trivial dynamical systems, the Poincar\'{e} map cannot be expressed by explicit analytical equations. Thus, obtaining information about the Poincar\'{e} map requires both solving the system of differential equations, and detecting when a point has returned to the Poincar\'{e} section. In \cite{ar13,hen,k19,k19a, par89,tuk,st15} authors discuss various numerical algorithms which fulfill both requirements.

Dynamics on the Poincar\'{e} map preserves many of the periodic and quasi-periodic orbits of the original system, and due to its dimensionally reduced form, it is often simpler to analyze than the original system. This is especially true for celestial mechanics which is perfectly suited for analysis via Poincar\'{e} maps \cite{poi} as Poincar\'{e} maps provide for a time scale separation by producing snapshots of dynamics on a sampling scale of the map period. In this note, we show that this can also be true for nonholonomic systems with periodic rheoneomic constraints. We have chosen the Bilimovich system as a sample toy-model generalizing the classical Suslov problem.

\section{Bilimovich system with periodic rheoneomic constraint}
Let us take Euler equations of motion
\[
\mathbb I \,\dot{\omega}=\mathbb I \,\omega\times \omega\, \qquad \omega=(p,q,r).
\]
where $\times$ denotes a vector product in $\mathbb R^3$,
and assume that the body frame is oriented in such a way that constraint $f=0$ has a fixed form (\ref{bil-c}).
In this frame inertia tensor $\mathbb I$ is a symmetric positive definite $3\times 3$ matrix.

We obtain equations of motion for the constrained system via the Lagrange D'Alembert principle that yields
\bq\label{eq-g}
\mathbb I\, \dot{\omega}=\mathbb I \,\omega\times \omega+\lambda\dfrac{\partial f}{\partial \omega}\,,\qquad
\dfrac{\partial f}{\partial \omega}=\left(
                   \begin{array}{c}
                    1 \\
                    -g(t) \\
                    0\\
                   \end{array}
                  \right)
\eq
where Lagrange multiplier $\lambda$ is determined on the condition that the constraint is satisfied
\[
\dfrac{df}{dt}=\left(\frac{\partial f}{\partial \omega}, \dot{\omega}\right)+\frac{\partial f}{\partial t}=0\,,
\]
Here $(x,y)$ denotes the scalar product of two vectors in $\mathbb R^3$. Using the equation of motion we get
\[
\lambda=-\frac{\left(\frac{\partial f}{\partial \omega},\mathbb I^{-1}(\mathbb I \,\omega\times \omega)\right)+\frac{\partial f}{\partial t}}{\left(\mathbb I^{-1}\frac{\partial f}{\partial \omega},\frac{\partial f}{\partial \omega}\right)}\,.
\]
Equations of motion (\ref{eq-g}) with this $\lambda$ preserve the quantity $p-g(t)q$. On the level set $p-g(t)q=\a$ we can substitute $p=\alpha+g(t)r$ into (\ref{eq-g}) and obtain a nonautonomous system of equations in the $(q,r)$ plane
\bq\label{eq-m2}
\dot{q}=Q(q,r,t)\qquad\mbox{and}\qquad \dot{r}=R(q,r,t)\,.
\eq
Any periodically forced nonautonomous differential equation can be converted to an au\-to\-no\-mous flow in a torus. To achieve this transformation,
simply introduce the third variable $s(t) = t$. The two-dimensional system of equations (\ref{eq-m2}) then becomes a three dimensional
autonomous system defined by equations
\bq\label{eq-a3}
\dot{q}=Q(q,r,s)\,,\qquad \dot{r}=R(q,r,s)\,,\qquad \dot{s}=1\,.
\eq
 Flow in this state space corresponds to the trajectory of flowing around a torus with a period $2\pi$. This naturally leads to a Poincar\'{e} section at $s=t_0$. The Poincar\'{e} map method is a standard tool to determine stability, chaos, and bifurcations for a periodically forced nonautonomous system of differential equations \cite{kuz,sh10}.

So, we take periodic functions $g(t)=g(t+T)$ and numerically solve the system of equations (\ref{eq-m2}) with random initial conditions $q(t_0)=q_0$ and $r(t_0)=r_0$. As a result, we get a set of points $x_k$ on $ (q, r) $-plane with coordinates
\[q_k=q(t_0+ kT)\qquad\mbox{and}\qquad r_k(t_0+ kT),\qquad k=0,\ldots,N\,.\]
The Poincar\'{e} map is a discrete map
\bq\label{poi-map}
\left(
 \begin{array}{c}
  q_{k} \\
  p_{k} \\
 \end{array}
\right)\to\left(
 \begin{array}{c}
  q_{k+1} \\
  p_{k+1} \\
 \end{array}
\right)\,,\quad\mbox{or}\qquad x_{k+1}= F(x_k)\,.
\eq
The study of autonomous systems usually begins with solving of equations
\bq\label{eq-qr}
Q(q,r,s)=0\,,\qquad R(q,r,s)=0\,,
\eq
with respect to $q$ and $r$ in order to get equilibrium points $(q_i^*,p_i^*)$ on the $(q,r)$-plane. In our case equations (\ref{eq-qr}) may have two, three, or four solutions depending on the parameter values. Different solutions allow us to explore the various Poincar\'{e} maps with different kinds of equilibrium points, attractors and strange attractors, compact and noncompact attractive curves, etc.

\subsection{Two critical points}
If $\mathbb I$ is a diagonal inertia tensor
\[\mathbb I_{12}=\mathbb I_{13}=\mathbb I_{23}=0\,,\]
 then equations (\ref{eq-m2}) are equal to
\[\begin{array}{rcl}
\dfrac{d q(t)}{d t}&=&Q(q,r,t)\equiv -\dfrac{1}{ g(t)^2 \mathbb I_{11} + \mathbb I_{22} }\times
\\
\\
& &
\Bigl(
\alpha(\mathbb I_{11} - \mathbb I_{33}) r(t)
 + g(t)(\mathbb I_{11} - \mathbb I_{22}) q(t) r(t)+ g(t)\frac{d g(t)}{dt}\mathbb I_{11} \,q(t)\Bigr)\,,
 \\
\\
\dfrac{d r(t)}{d t}& =&R(q,r,t)\equiv \dfrac{ (\mathbb I_{11} - \mathbb I_{22})\bigl(g(t) q(t) + \alpha\bigr) q(t)}{\mathbb I_{33}}\,.
\end{array}
\]
If $g(t)=g(-t)$, the equations are invariant with respect to involution
\[
t\to -t,\qquad q\to q\,,\qquad r\to -r\,.
\]
 So, trajectories in the manifold of states of motion are grouped in symmetric pairs and, therefore, attractors and repellers on the Poincar\'{e} maps meet in pairs which are symmetrical to the replacement.

Two solutions of the corresponding system of equations (\ref{eq-qr}) have the following form
\bq\label{sol-1}
x^*_1=(0,0)\qquad \mbox{and}\qquad
x^*_2=\left(-\frac{\alpha}{g(t)},
\frac{\mathbb I_{11}}{(\mathbb I_{22} -\mathbb I_{3 3})}\frac{d g(t)}{dt}
\right)\,.
\eq
These solutions are the counterparts of equilibrium points for the Poincar\'{e} maps for the periodically forced nonautonomous system (\ref{eq-m2}).

Below we introduce $g(t)=\cos t$ and $t_0=0$, so that
\bq\label{case1-r}
x^*_1=(0,0)\,,\qquad\mbox{and}\qquad x^*_2=\left(-\frac{\alpha}{\cos t},\frac{\mathbb I_{11}}{(\mathbb I_{22} -\mathbb I_{3 3})}\sin(t))\right),
\eq
and present the Poincar\'{e} section data obtained by numerical integration of (\ref{eq-m2}).

In generic cases at
\[
\mathbb I_{11}<\mathbb I_{22}<\mathbb I_{33}\qquad \mbox{and}\qquad
\mathbb I_{11}>\mathbb I_{22}>\mathbb I_{33}
\]
the Poincar\'{e} sections haves noncompact attractive curves in comparison to other cases when all the trajectories are compact.
Let us give a few Poincar\'{e} sections for equations (\ref{eq-m2}) with noncompact attractive curves at
\[
A:\, \mathbb I=diag(1,2,4)\,,\quad B:\,\mathbb I=diag(4,2,1)
\]
and with compact attractive curves at
\[
C:\, \mathbb I=diag(1,4,2)\,,\quad D:\,\mathbb I=diag(4,1,2)\,.
\]
\begin{figure}[H]
\center{\includegraphics [width=0.4\linewidth]{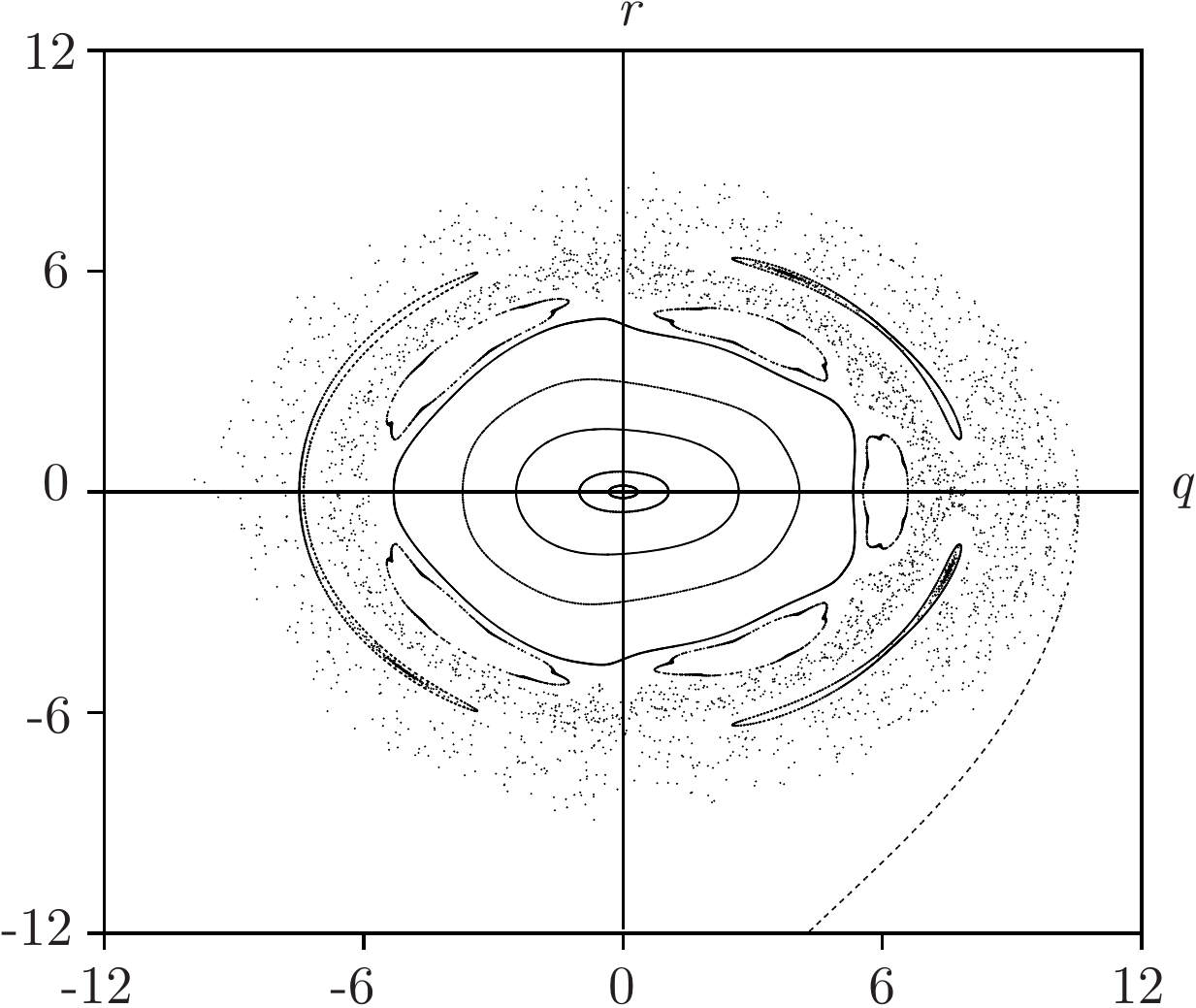} A}\\
\center{\includegraphics [width=0.4\linewidth]{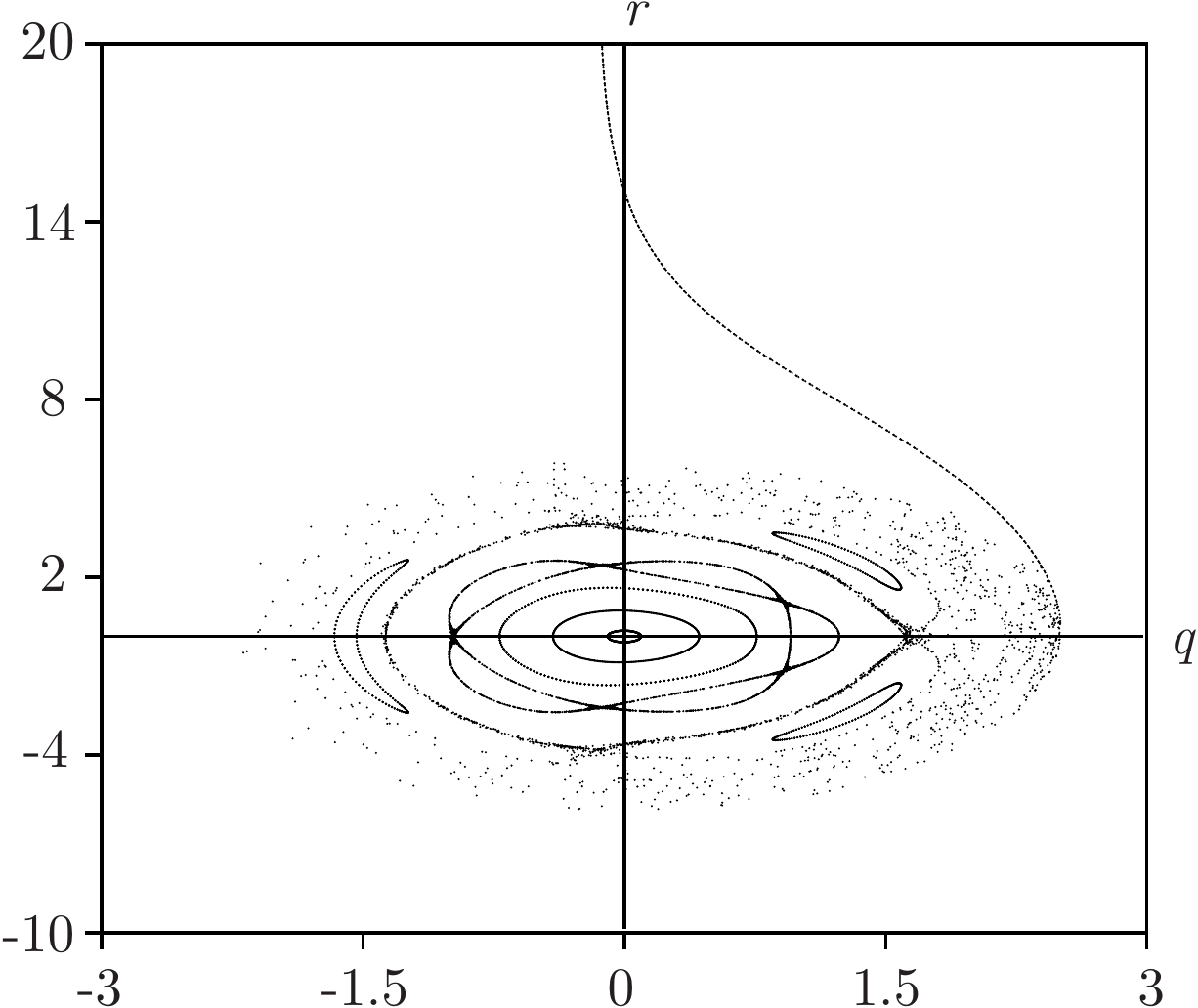} B}\\
\center{\includegraphics [width=0.4\linewidth]{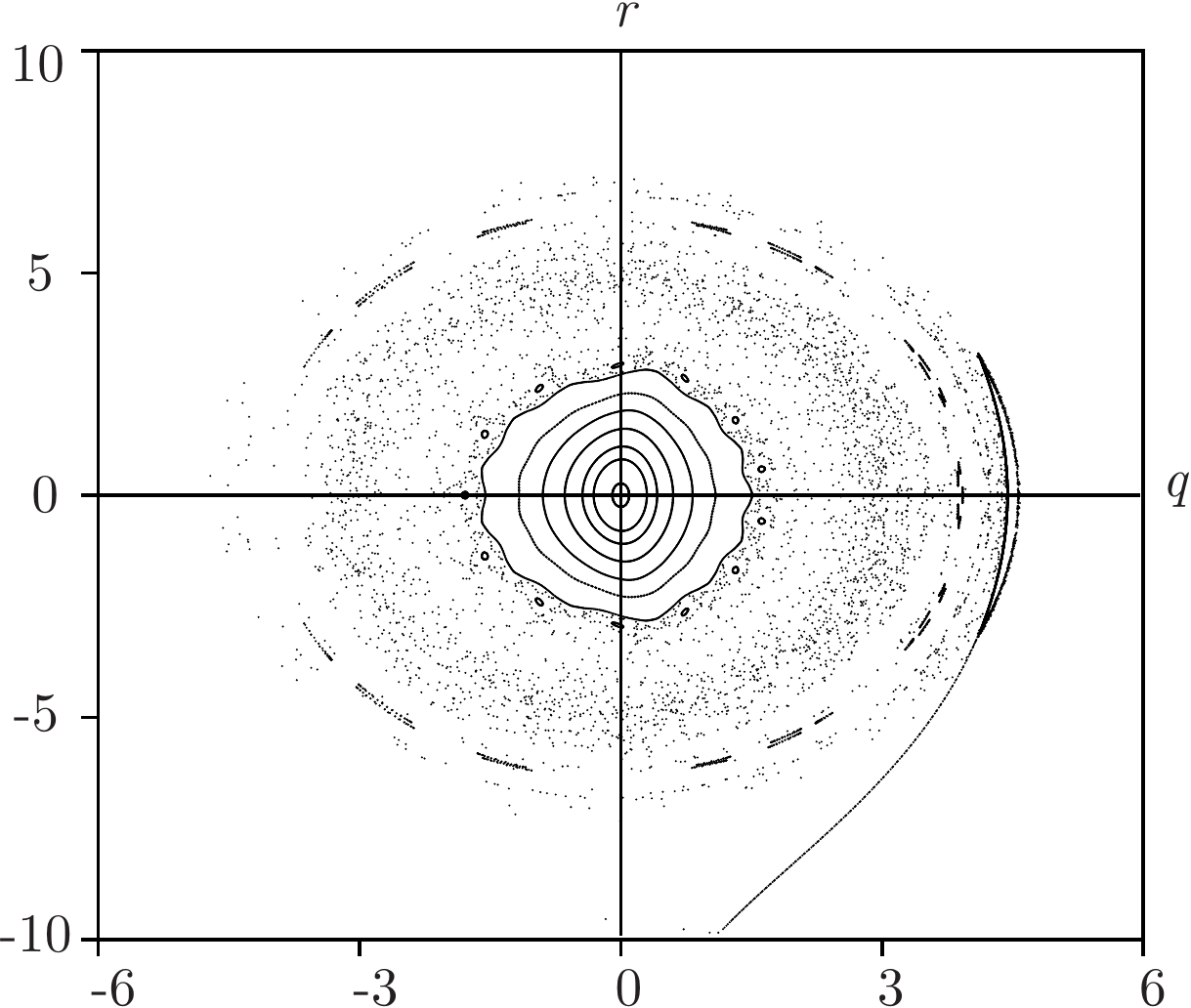} C}\\
\center{\includegraphics [width=0.4\linewidth]{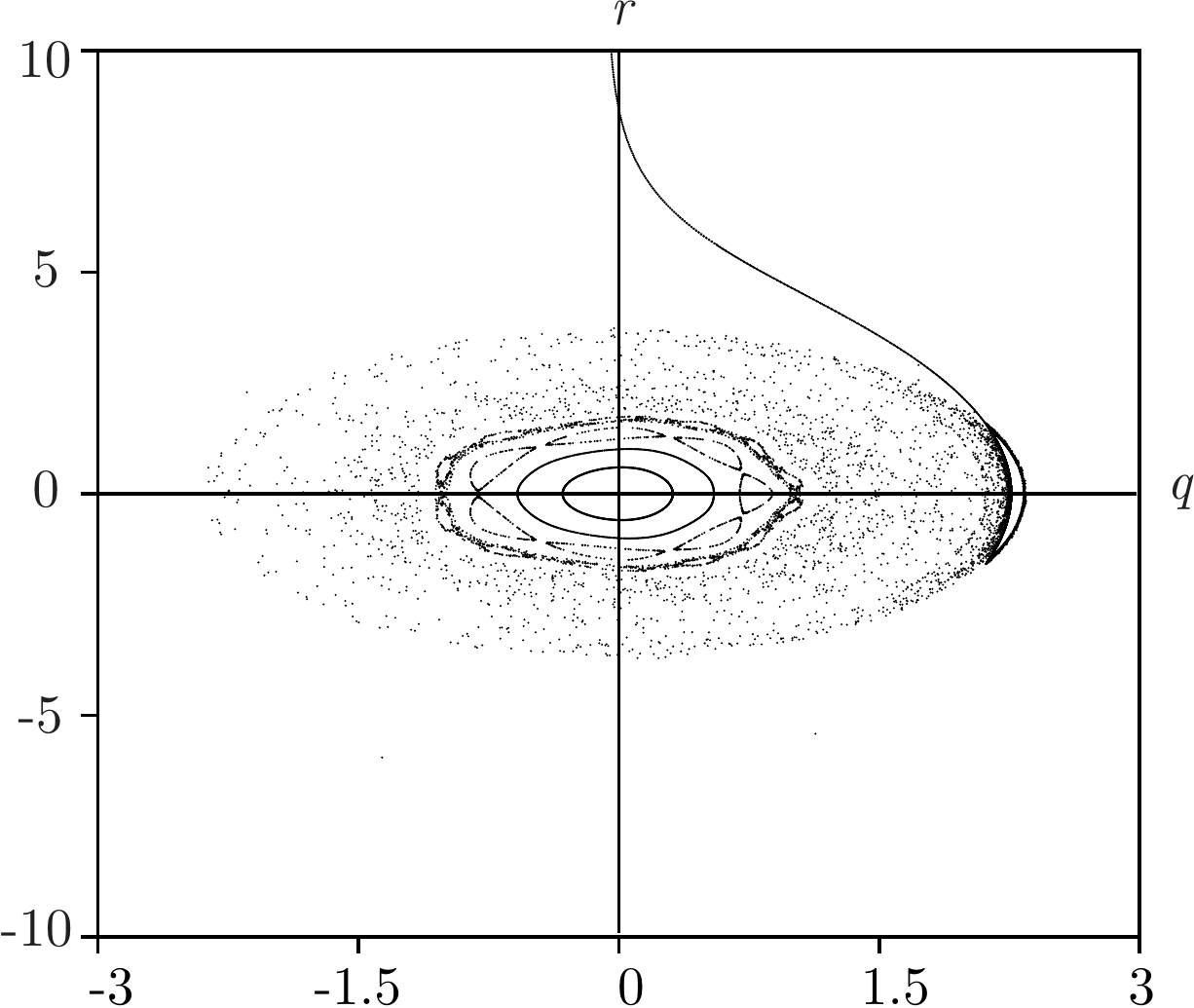} D}
\caption{Poincar\'{e} maps with compact and noncompact attractive curves at  $\alpha=0.1$}
\label{ris1}
\end{figure}
The Poincar\'{e} sections in Fig.\ref{ris1} A-B consist of invariant curves associated with the quasi-periodic solutions of  (\ref{eq-m2}), i.e. the so-called islands of critical tori, and chaotic trajectories turning into a noncompact attractive curve. In Fig.\ref{ris1} C-D we can also find invariant curves associated with the quasi-periodic solutions of (\ref{eq-m2}) and an attractive invariant curve corresponding to a crescent moon-shaped torus in phase space.

\begin{figure}[H]
\center{\includegraphics [width=0.5\linewidth]{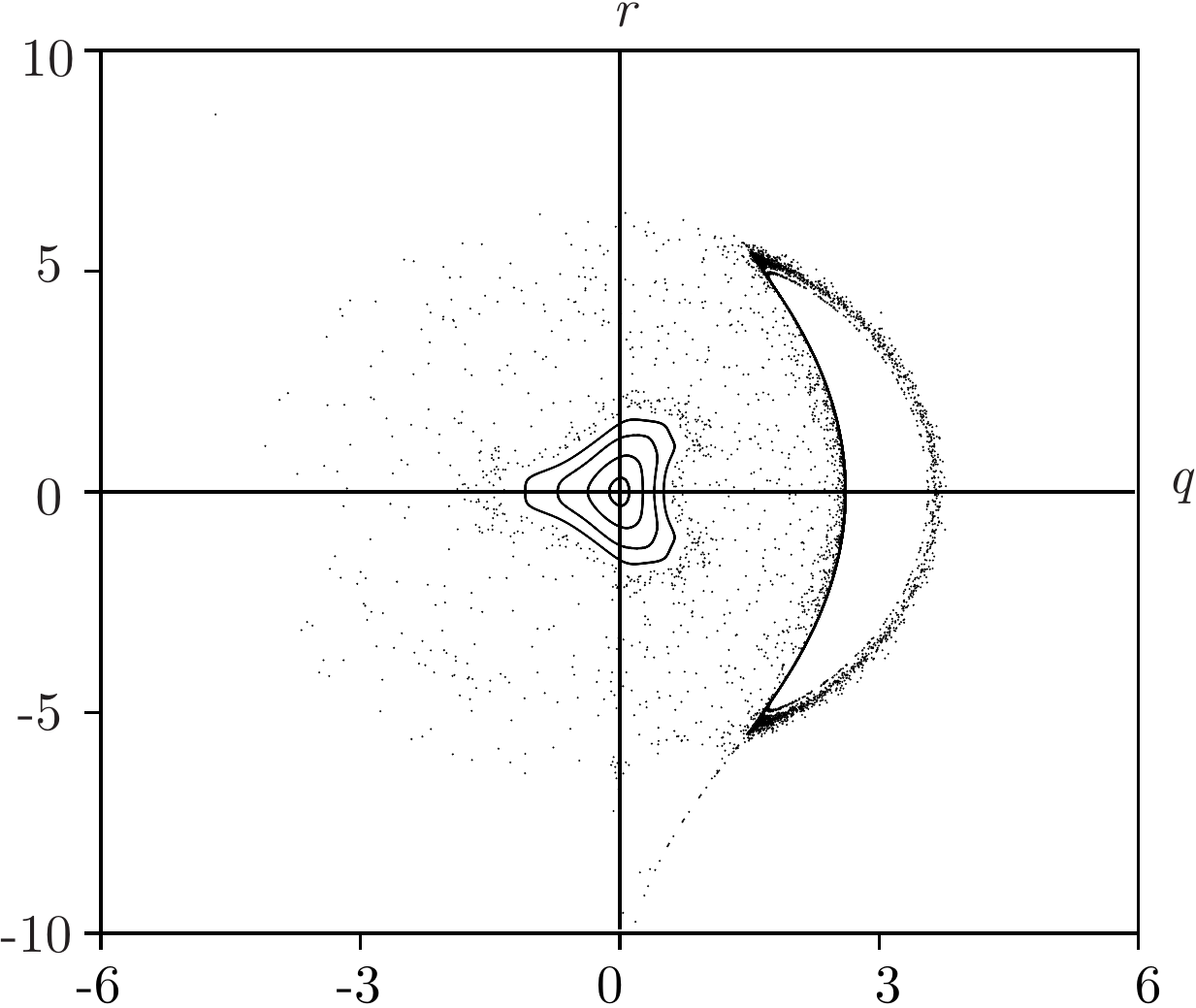} C}\\
\center{\includegraphics [width=0.5\linewidth]{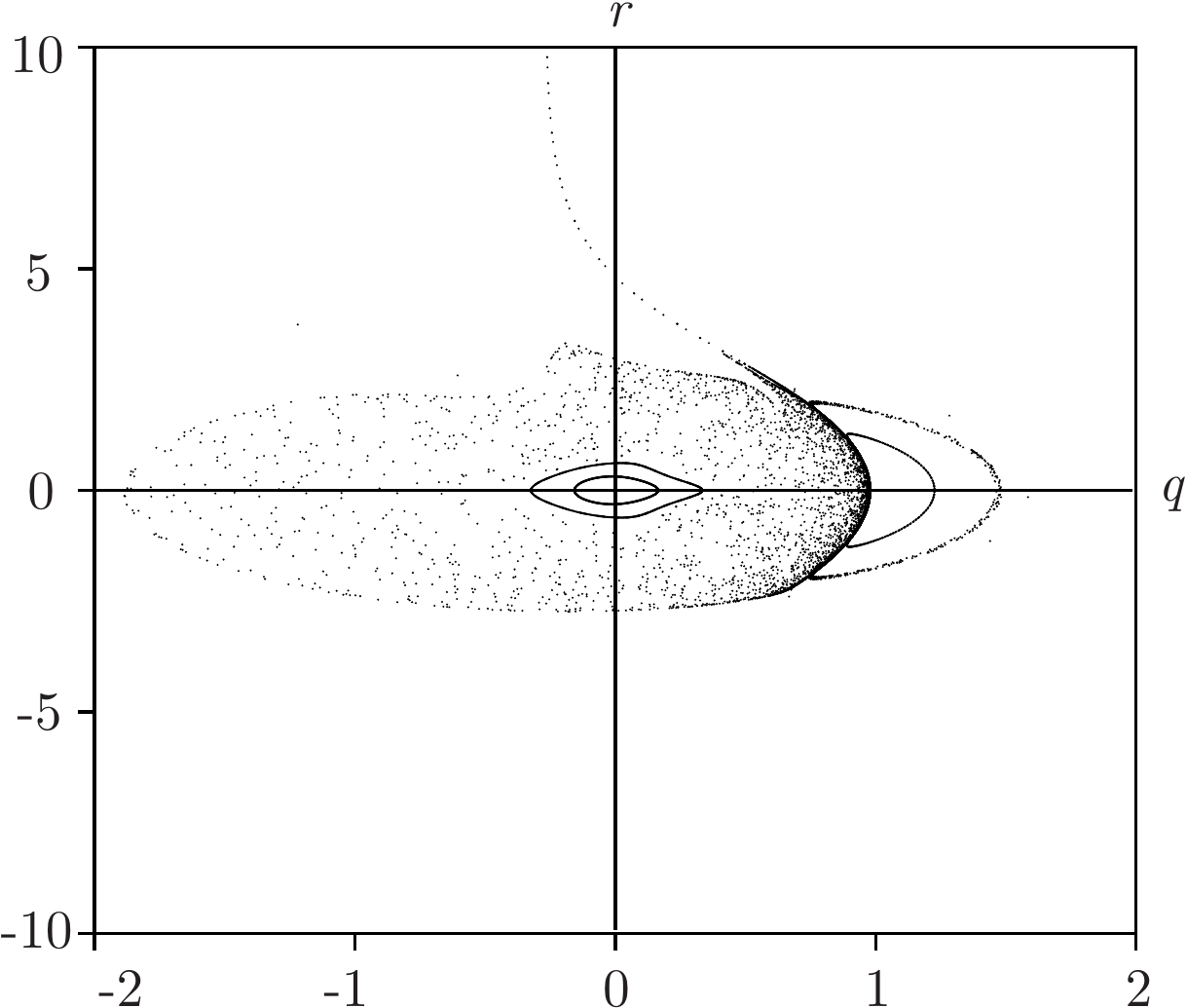} D}
\caption{Poincar\'{e} maps with crescent moon-shaped curves at $\alpha=0.4$}
\label{ris2}
\end{figure}

The crescent moon-shaped curves have merged to form a single island of tori in phase space at $\alpha=0.4$ which is shown in Fig.\ref{ris2}
\par\noindent
 The many points existing outside of these two islands are located in a region passed through by the unstable aperiodic orbits in this section $t_0=0$.

If the time derivative of kinetic energy
 \[
 T'=\dfrac{dT}{dt}=\alpha\lambda=-\alpha\frac{\left(\frac{\partial f}{\partial \omega},\mathbb I^{-1}(\mathbb I \,\omega\times \omega)\right)+\frac{\partial f}{\partial t}}{\left(\mathbb I^{-1}\frac{\partial f}{\partial \omega},\frac{\partial f}{\partial \omega}\right)}
 \]
 in the given region of phase space is always greater than zero, then the rigid body starts to rotate faster. If $T'$ is always less than zero, then this rotation attenuates. If $T'$ is alternating, then trajectories of the rigid body in velocity space will be compact.

\begin{figure}[H]
\center{\includegraphics [width=0.5\linewidth]{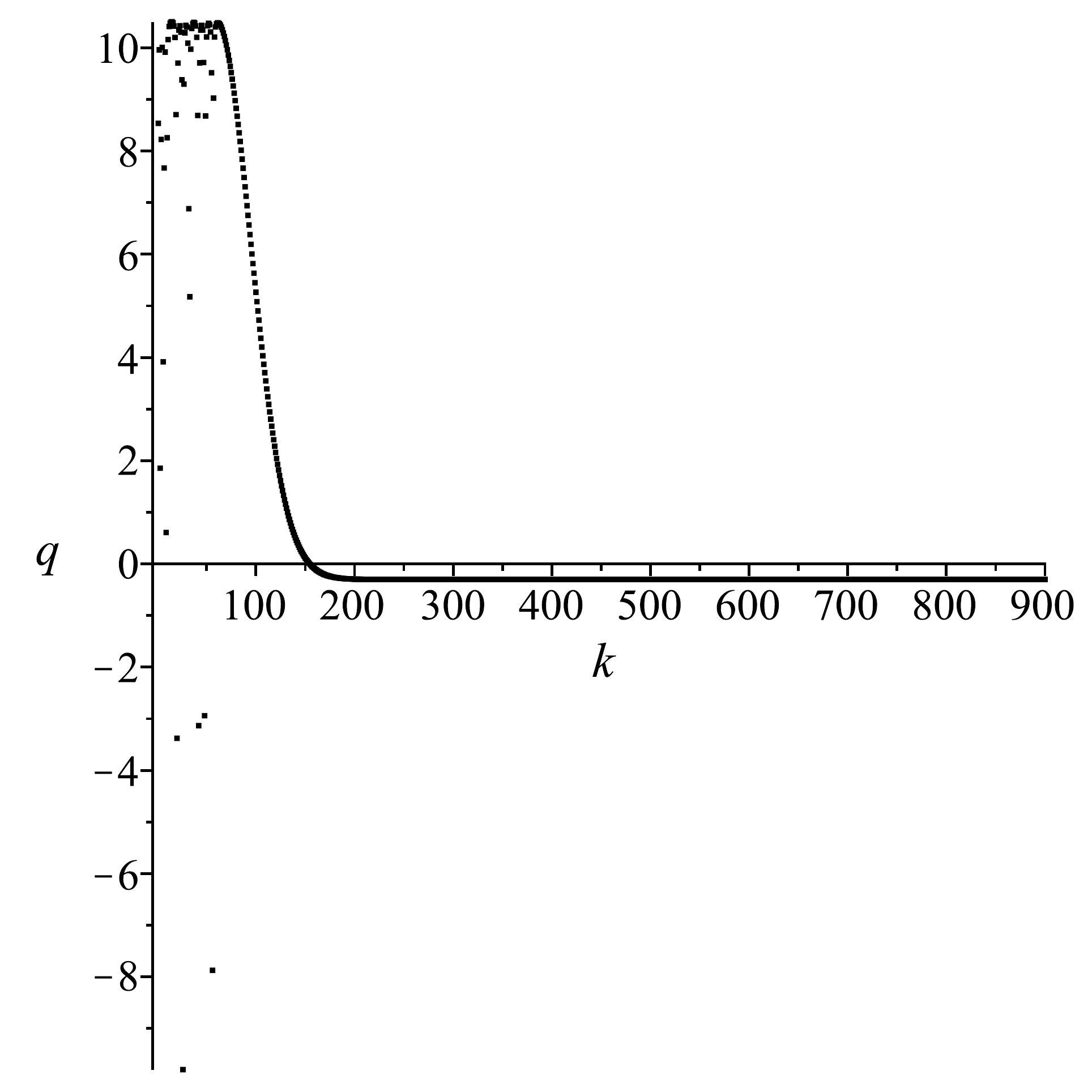}}\\
\center{\includegraphics [width=0.5\linewidth]{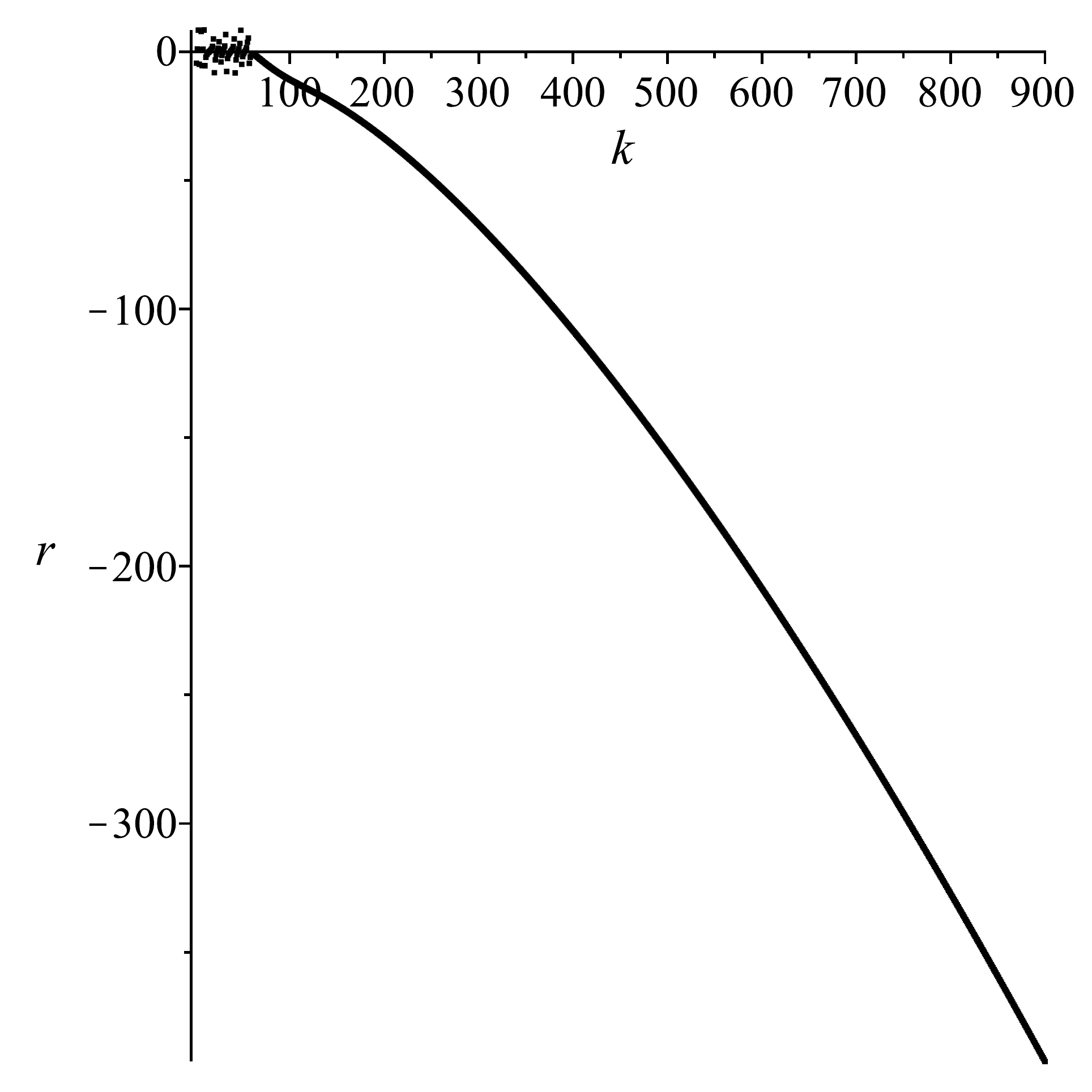}}
\caption{Poincar\'{e} section data $q_k$ and $r_k$ at $\alpha=0.1$}
\label{ris3}
\end{figure}

It is rather cumbersome to exactly estimate derivative $T'$ in the given region of phase space by using the Poincar\'{e} map method. For instance, let us consider data from ten random trajectories turning into a noncompact attractive curve in Fig.\ref{ris1}A. Results of numerical integration of (\ref{eq-m2}) with initial conditions $q_0=0$ and $r_0=8$ are given in Fig.\ref{ris3}.
\par\noindent
At $k=200\ldots 900$ by minimizing the least-squares error we can fit a univariate polynomial to data $q_k$ and $r_k$ .
By averaging coefficients of ten such polynomials associated with random initial conditions we obtain
 \[\begin{array}{rcl}
q(t)&=&-0.300263\,,\\ \\ r(t)&=& - 0.0002545\,t^2 - 0.2386278\,t+26.779718=at^2+bt+c\,.
\end{array}
\]
Thus, the Poincar\'{e} map in this region reads as
\[
q_{k+1}=q_k\,,\qquad
 r_{k+1}= r_k + 2\pi\sqrt{b^2-4ac + 4ar_k\,} + 4a\pi^2\,.
\]
 Although this mapping is not an exact representation of (\ref{poi-map}), it allows us to describe the qualitative dynamics of the rigid body under Bilimovich's constraint (\ref{bil-c}). Indeed, we can say that on average in this region of phase space the first and second components of angular velocity are constants, whereas the absolute value of the third component in increasing.

At $\mathbb I_{13}=\mathbb I_{23}=0$ equations (\ref{eq-m2}) also have only two solutions. According to \cite{bt20} in this case we can integrate the nonautonomous equations (\ref{eq-m2}) in terms of hyperbolic functions at $\alpha=0$.

\subsection{Three critical points}
At $\mathbb I_{12}=\mathbb I_{23}=0$, equations (\ref{eq-qr}) have two fixed solutions
\[
x^*_1=\left( 0,\frac{\mathbb I_{33}-\mathbb I_{11} \pm \sqrt{(\mathbb I_{11} -\mathbb I_{33})^2 + 4\mathbb I_{13}^2 \,}}{2\mathbb I_{13}}\,\alpha\right)
\]
and one solution depending on time. If
\[
\mathbb I=\left(
      \begin{array}{ccc}
       1 & 0 & 0.1 \\
       0 & 2 & 0 \\
       0.1 & 0 & 4 \\
            \end{array}
     \right)\,,\qquad g(t)=\cos t\,,
\]
 the Poincar\'{e} transversal plane at $t_0=0$ looks like
\begin{figure}[H]
\center{\includegraphics [width=0.5\linewidth]{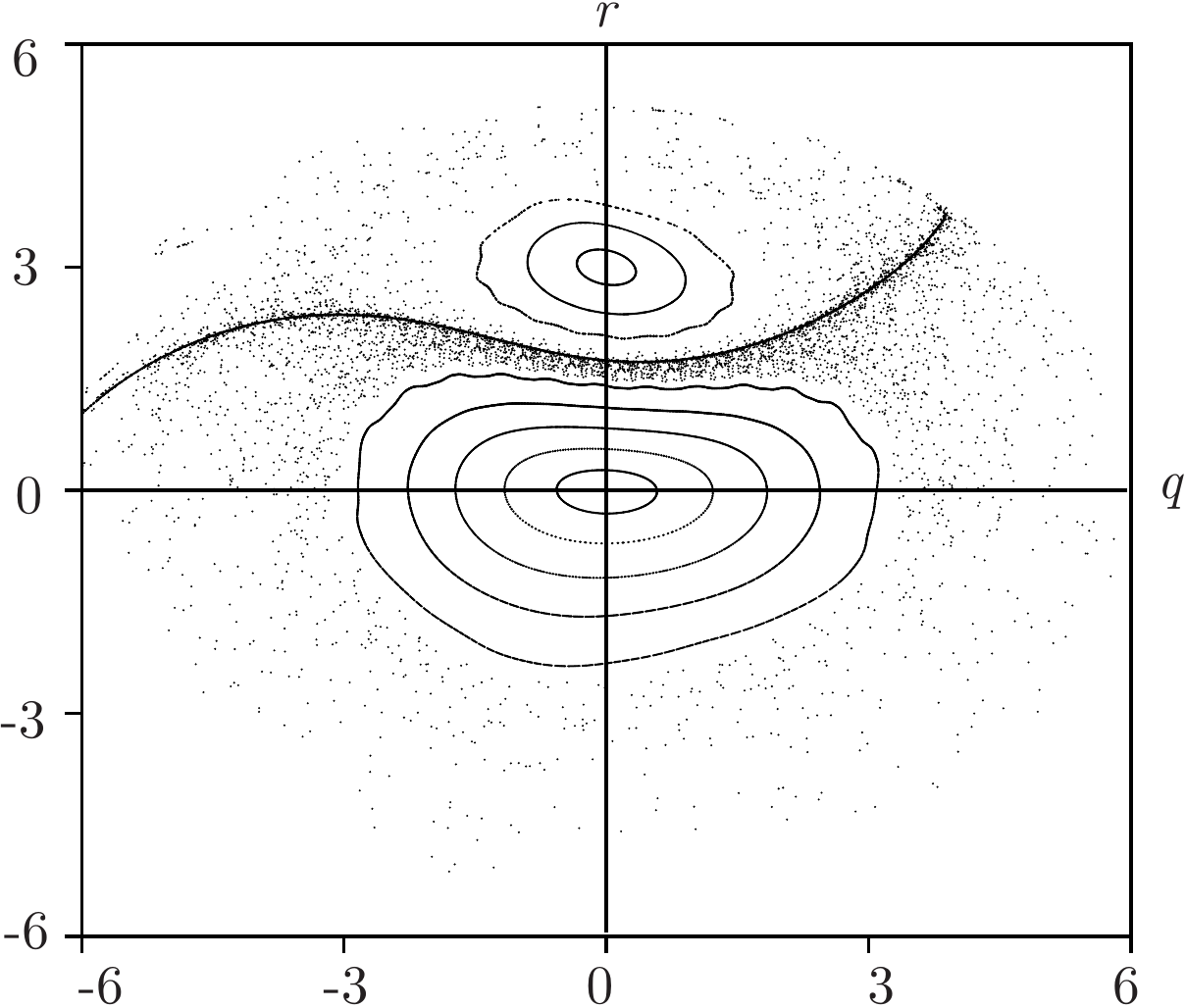}A}
\center{\includegraphics [width=0.5\linewidth]{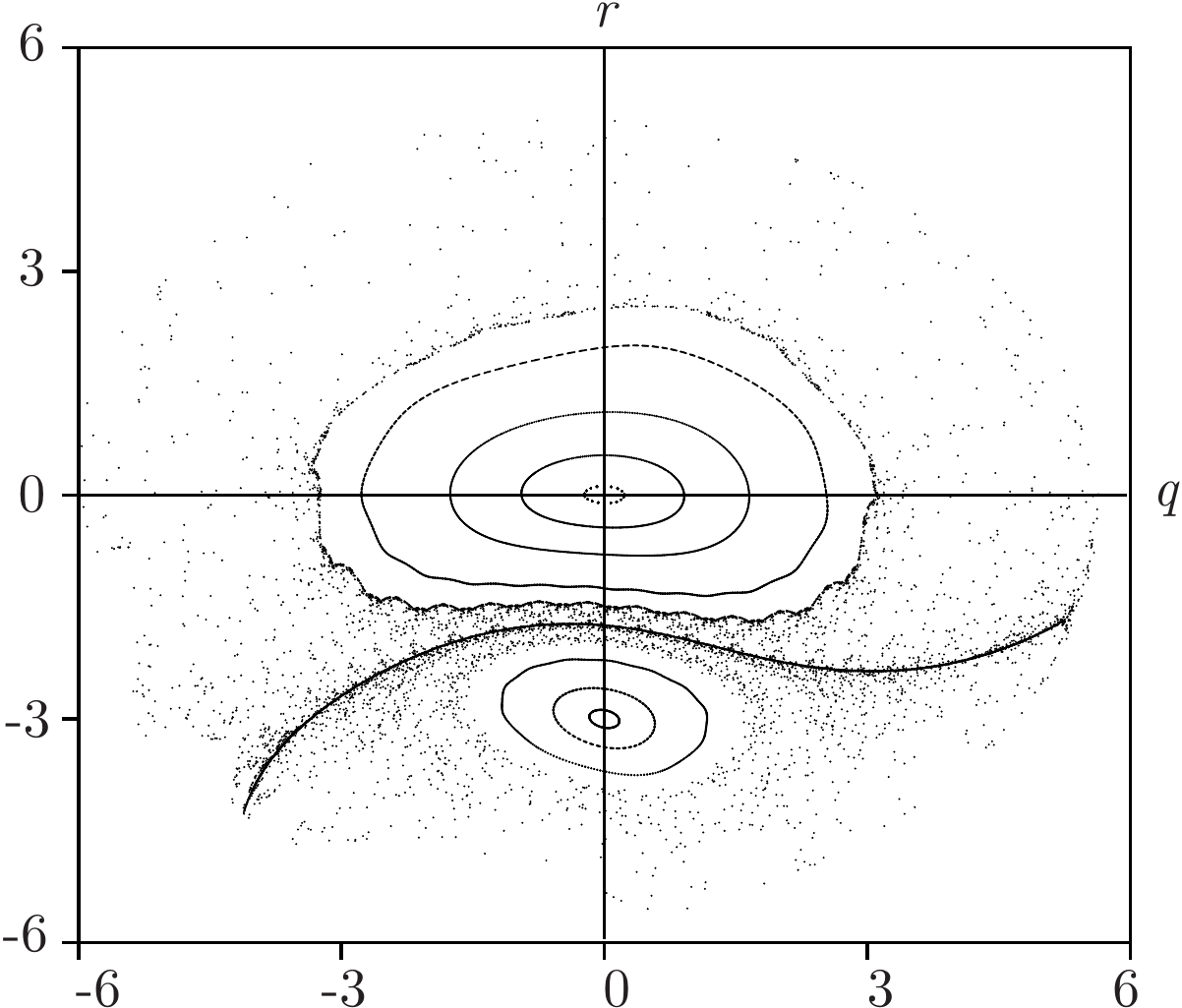}B}
\caption{Poincar\'{e} maps at $\alpha=\pm 0.1$}
\label{ris4}
\end{figure}
\par\noindent
 In Fig.\ref{ris4}A-B we can see quasi attractors for the system of nonautonomous equations (\ref{eq-a3}). These
quasi attractors have the following Lyapunov exponents
 \[\Lambda_1=0.022, \quad \Lambda_2=-0.034,\quad \Lambda_3=0,\qquad \sum_{i=1}^3\Lambda_i=-0.012\,,\]
and Lyapunov dimension
\[
D=1+\dfrac{\Lambda_1}{|\Lambda_2|}=1.647.
\]
Values of Lyapunov exponents $\Lambda_i$ and dimension $D$ are independent of the sign of parameter $\alpha$.

\begin{figure}[H]
\center{\includegraphics [width=0.5\linewidth]{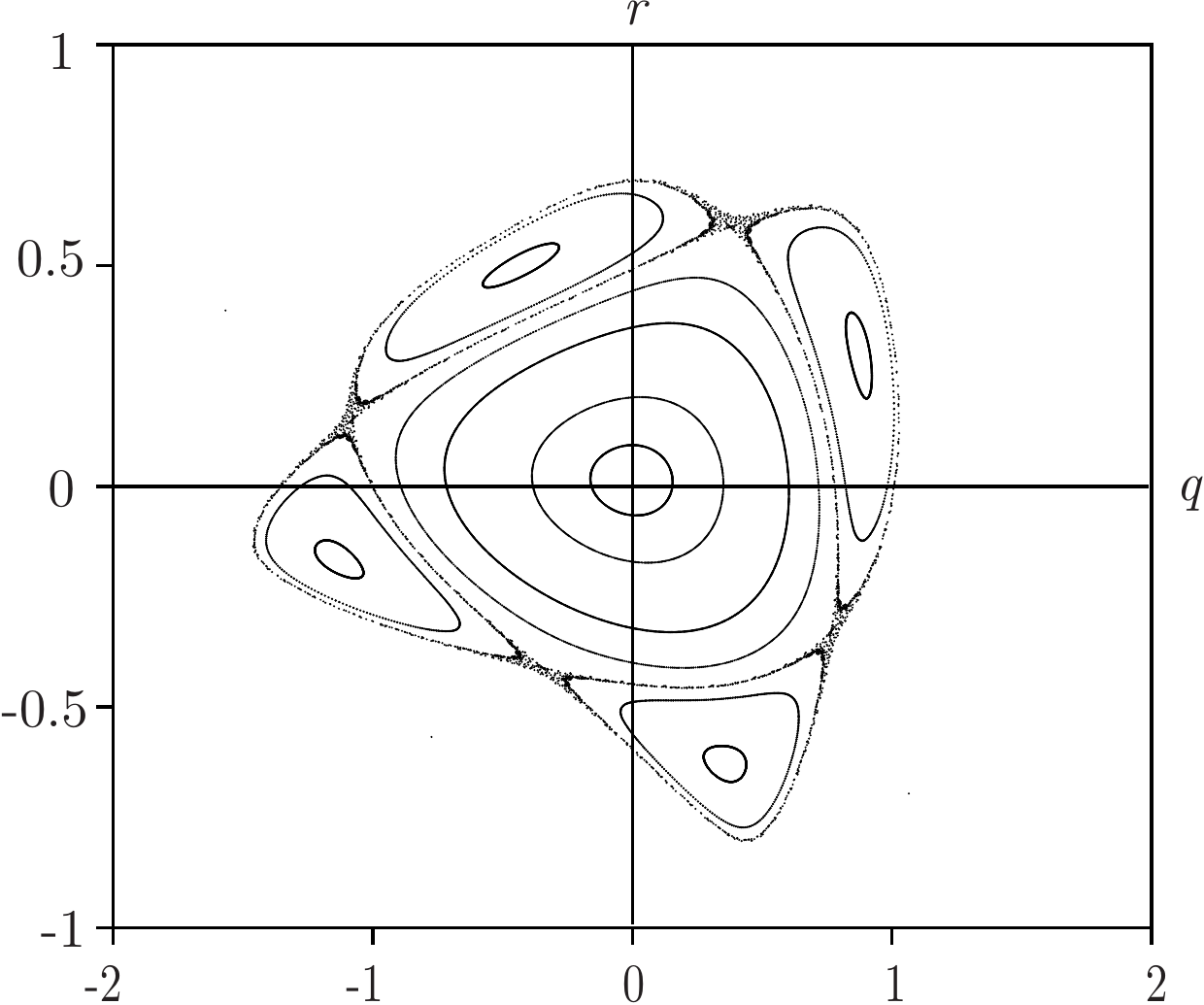}}\\
\center{\includegraphics [width=0.5\linewidth]{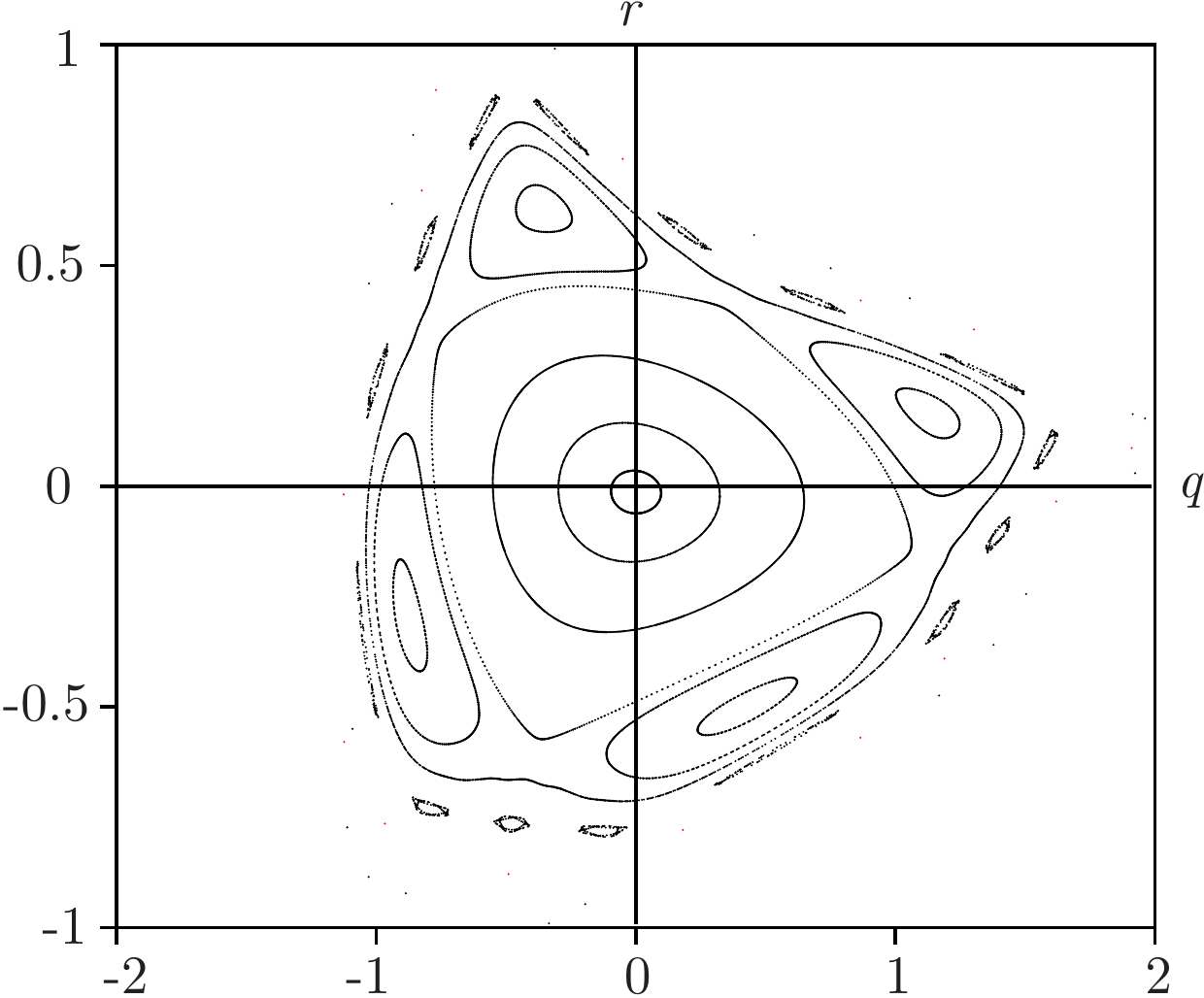}}\\
\caption{Central part of the Poincar\'{e} sections at $\alpha=\mp0.4$}
\label{ris5}
\end{figure}

\begin{figure}[H]
\center{\includegraphics [width=0.5\linewidth]{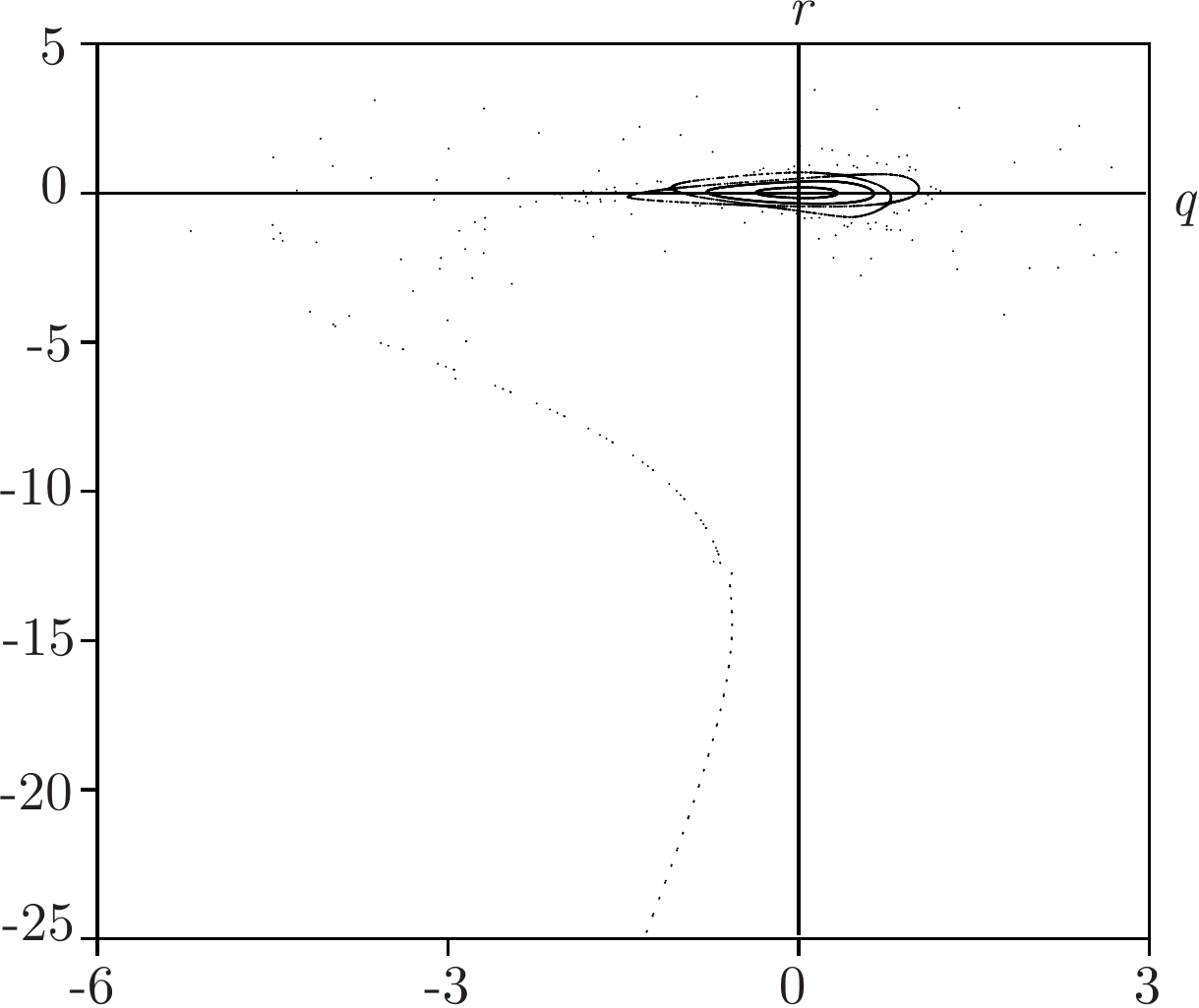}A}\\
\center{\includegraphics [width=0.5\linewidth]{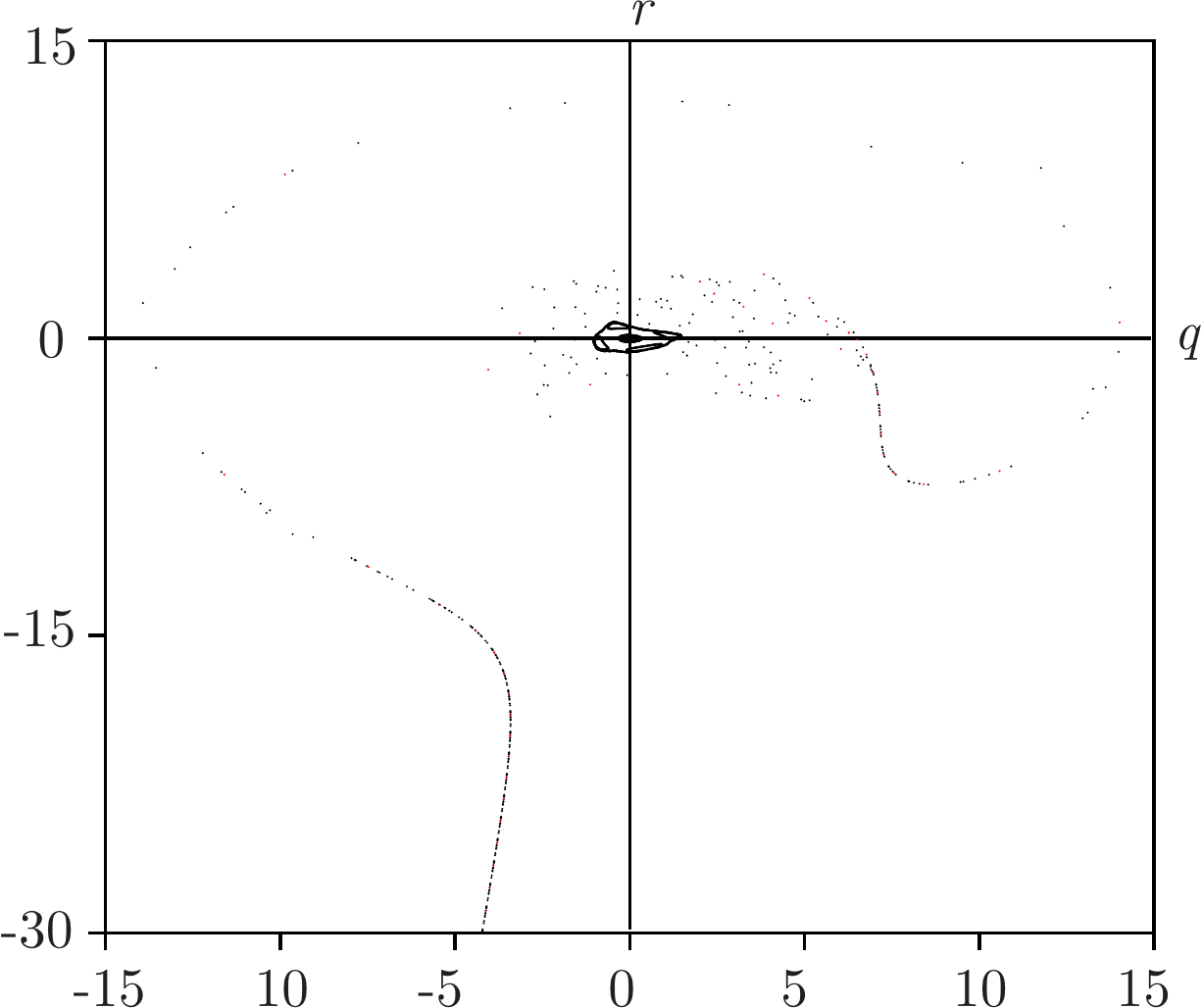}B}
\caption{Poincar\'{e} sections with noncompact invariant curves at $\alpha=\mp0.4$}
\label{ris6}
\end{figure}

 This quasi attractor disappears upon an increase of $|\alpha|$, and we get the islands of invariant curves associated with the quasiperiodic solutions and noncompact attractive curve at $\alpha=\pm0.4$, see Fig.\ref{ris5} and Fig.\ref{ris6}.
\par\noindent
\par\noindent
At $\alpha=-0.4$,
polynomials describing the Poincar\'{e} data at $k=200\ldots 900$ are equal to
 \[
q(t)=-0.5241223 - 0.1461676\,t\,,
\qquad
r(t)=0.00009\,t^2 - 1.5651429\,t+7.4115722.
\]
after averaging by ten trajectories with random initial conditions, see Fig.\ref{ris6}A.
At $\alpha=0.4$ these polynomials have the following form
 \[
q(t)=-2.7790128 - 0.14597897\,t\,,
\qquad
r(t)=0.00009 t^2- 1.5598174\,t +7.966923.
\]
These quadratic polynomials $r(t)$ can only be used locally and bear little resemblance to the true map (\ref{poi-map}) at $k\to \infty$ due to leading coefficients being positive.

To get an asymptotic we can fit this Poincar\'{e} section data by linear polynomials
\[\alpha=-0.4,\quad r(t)=- 1.4616911\,t-17.1864607\]
and
\[
\alpha=0.4\quad  r(t)=- 1.459806\,t-15.8130501
\]
or by cubic polynomials $r(t)=-at^3+bt^2+ct+d$ with negative leading coefficient $a\sim 5\cdot 10^{-8}$. Another way for achieving this is numerical integration of the equations of motion over a larger interval 	$[0,2\pi k]$, $k>900$ without precision loss.

In both cases, we can say that absolute values of angular velocity components increase in proportion to each other.

\subsection{Four critical points}
At $\mathbb I_{13}\neq 0$ and $\mathbb I_{23}\neq 00$, equations (\ref{eq-qr}) have four solutions
Let us present a gallery of the Poincar\'{e} sections for the following inertia tensor
\begin{equation}\label{i3}
\mathbb I=\left(
      \begin{array}{ccc}
       1 & 0 & 0.5 \\
       0 & 2 & 0.3 \\
       0.5 & 0.3 & 4 \\
            \end{array}
     \right)\,,
\end{equation}
which is designed to illustrate the origin and destruction of a strange attractor.

\begin{figure}[H]
\center{\includegraphics [width=0.5\linewidth]{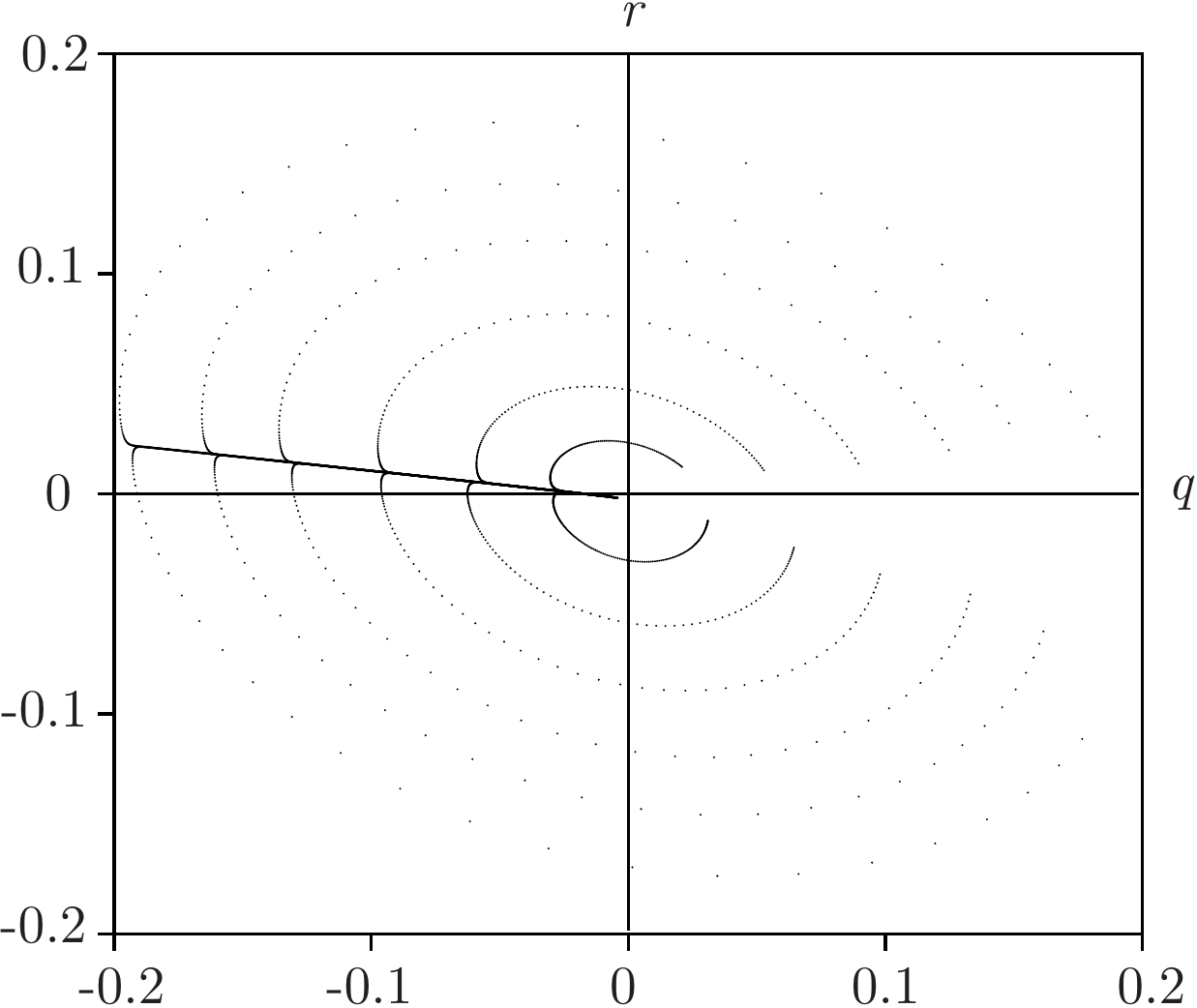}A}\\
\center{\includegraphics [width=0.5\linewidth]{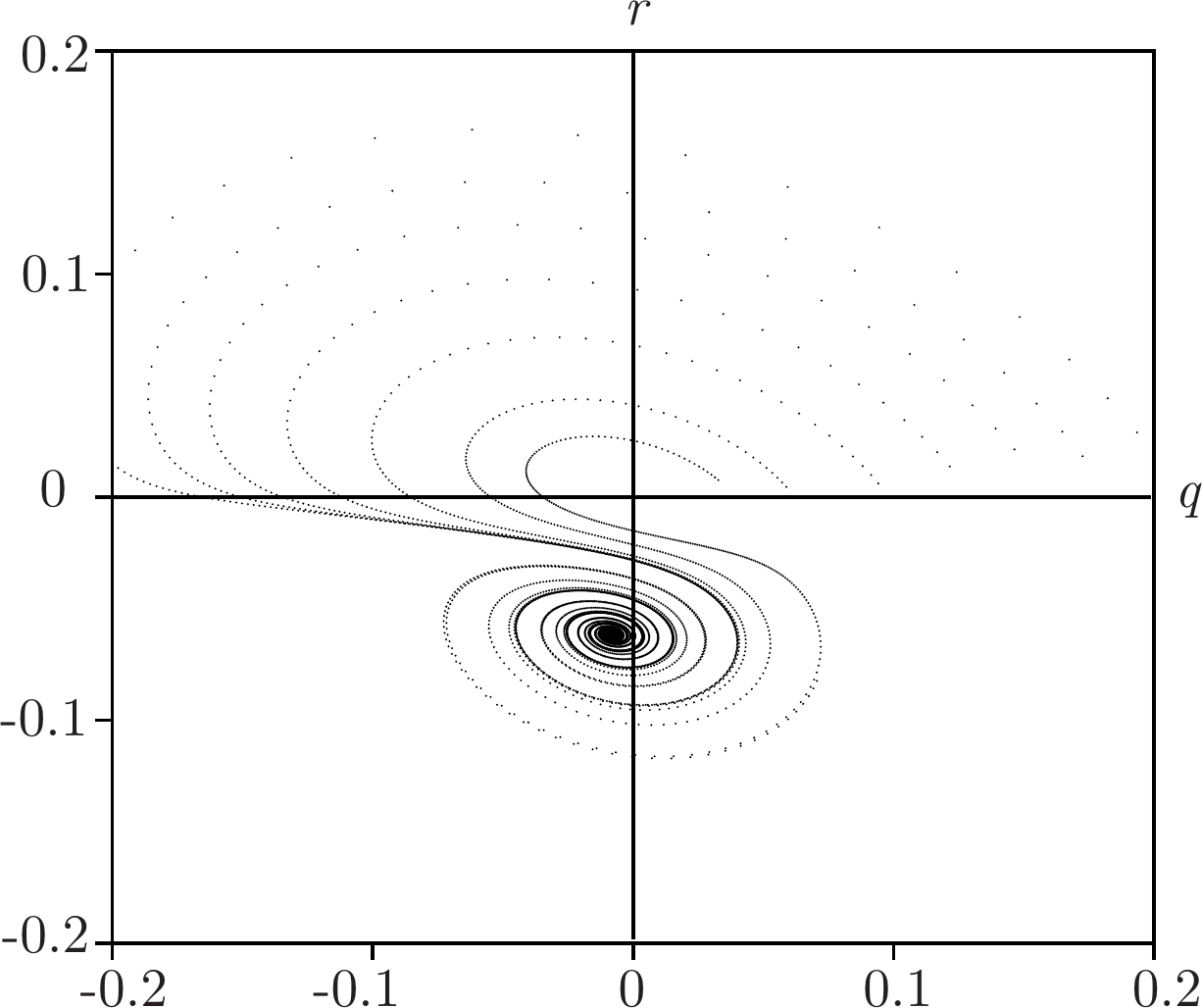}B}
\center{\includegraphics [width=0.5\linewidth]{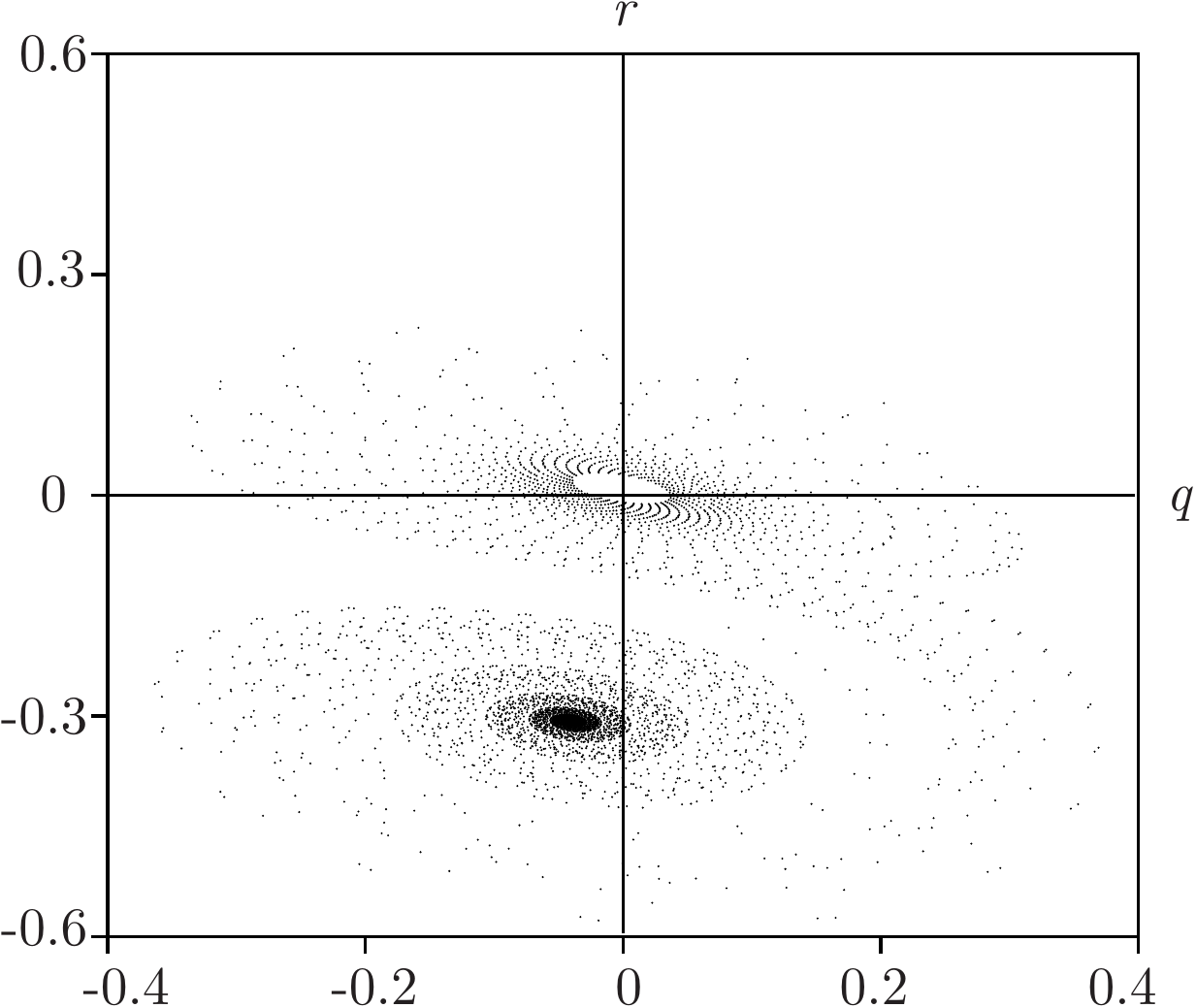}C}
\caption{Poincar\'{e} sections at $\alpha=-0.001$, $\alpha=-0.01$ and $\alpha=-0.05$}
\label{ris7}
\end{figure}

In the first stage, periodic trajectories become quasiperiodic trajectories attracted to the invariant curve in Fig.\ref{ris7}A, which later forms an unstable focus in Fig.\ref{ris7}B. Then trajectories form a stable focus and an unstable focus or repeller in Fig.\ref{ris7}C.
\par
\begin{figure}[H]
\center{\includegraphics [width=0.5\linewidth]{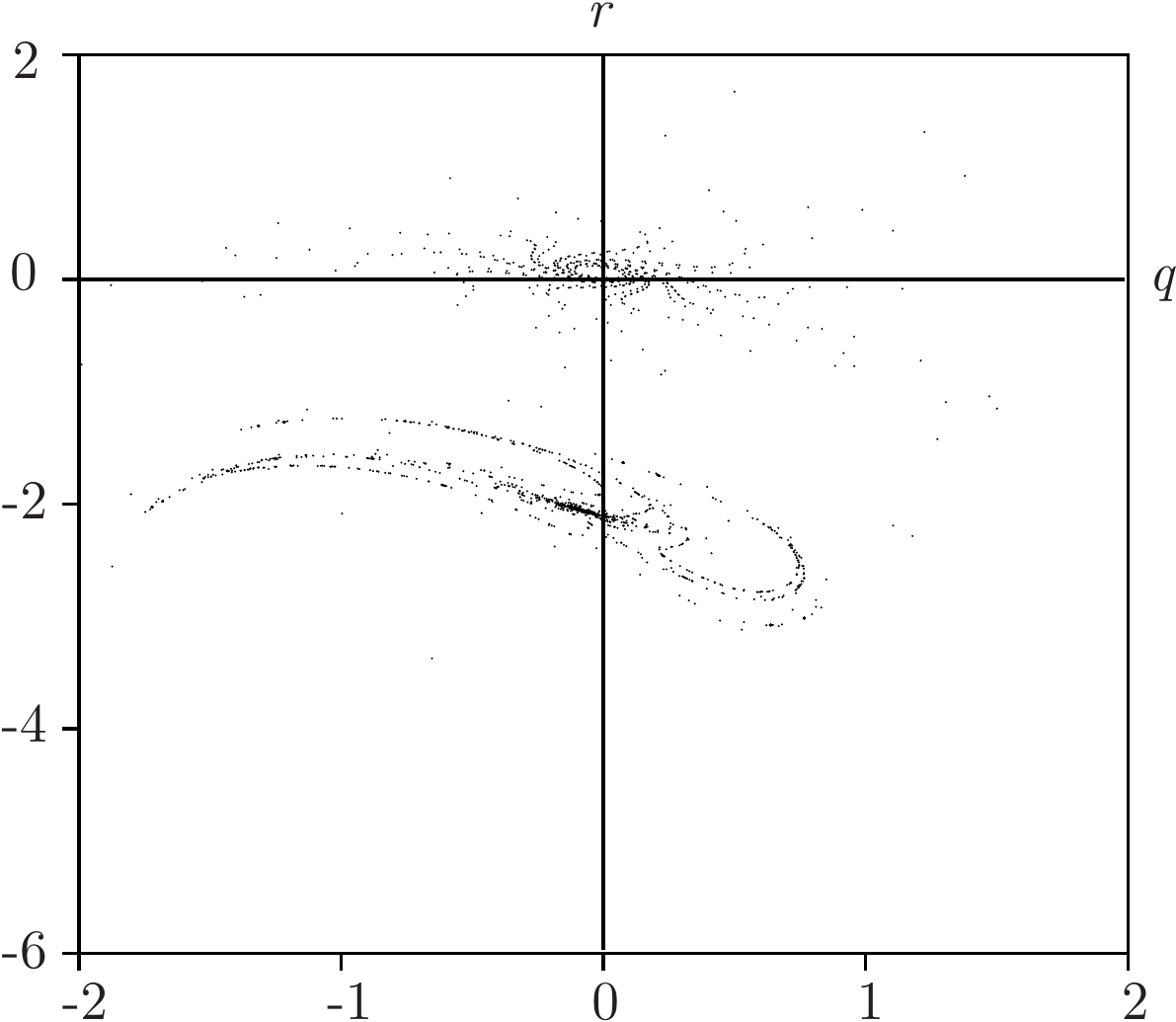}A}
\center{\includegraphics [width=0.5\linewidth]{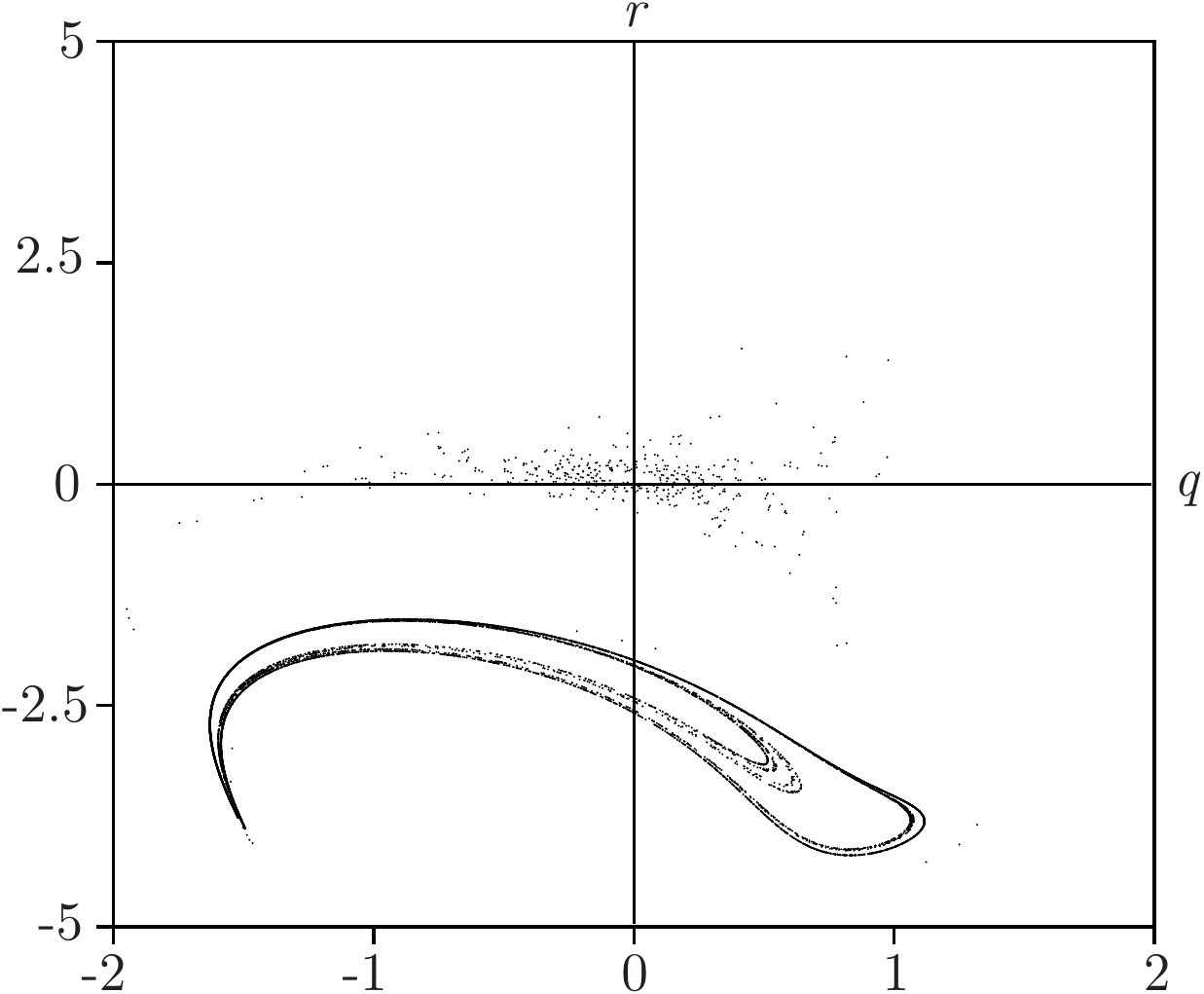}B}
\center{\includegraphics [width=0.5\linewidth]{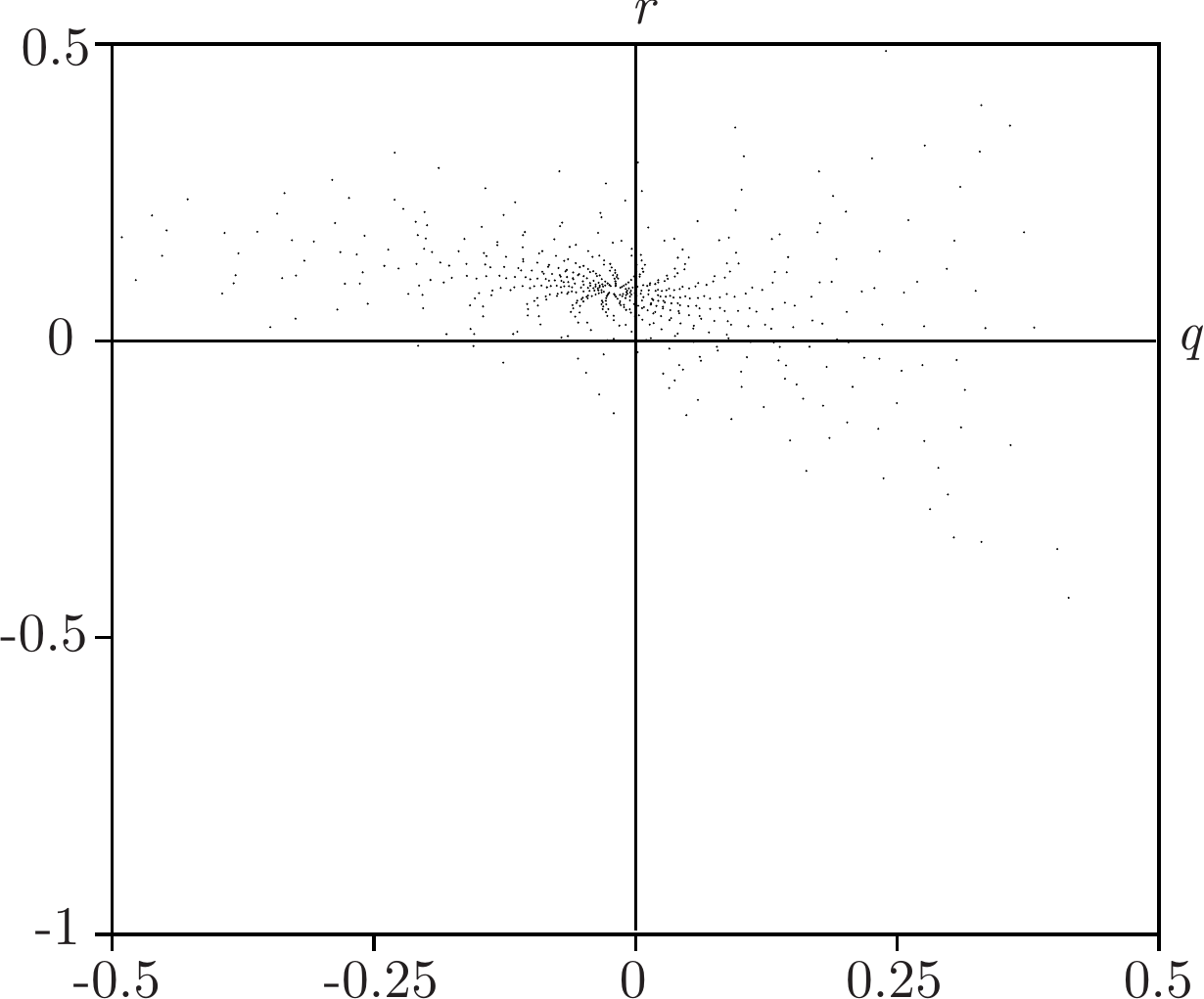}C}
\caption{Poincar\'{e} sections at $\alpha=-0.37$, $\alpha=-0.45$ and $\alpha=-0.5$}
\label{ris8}
\end{figure}

\noindent
In the second stage, in Fig.\ref{ris8}A-B the unstable focus goes into the strange attractor, and then the strange attractor disappears in Fig\ref{ris8}C.
At $\alpha=-0.5..-0.37$ rotation near a stable focus changes direction from clockwise to counterclockwise and vise versa.
\par\noindent
 Lyapunov exponents for the strange attractor in Fig.\ref{ris8}B are equal to
\[
\Lambda_1\simeq 0.082\,,\qquad
\Lambda_2\simeq -0.385\,,\qquad
\Lambda_3\simeq 0\,,
\]
so its Lyapunov dimension is
\[
D=1+\dfrac{\Lambda_1}{|\Lambda_2|}=1.21\,.
\]
It is compatible with fractal dimension of the strange attractor for H\'{e}non map $D=1.26$ or Lorentz map $D=2.07$.
The corresponding bifurcation diagram looks like
 \begin{figure}[H]
\center{\includegraphics [width=0.5\linewidth]{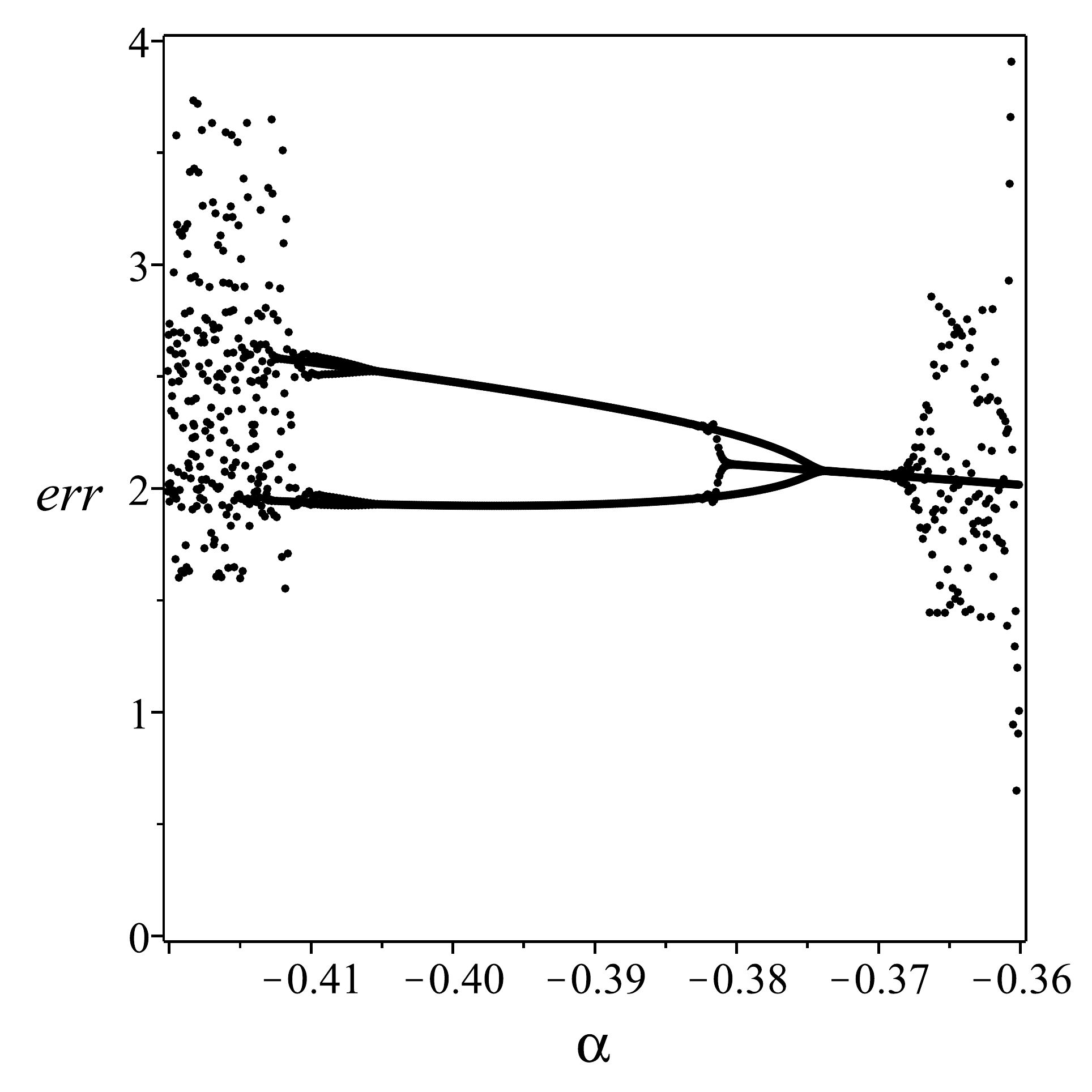}}
\caption{Bifurcation diagram at $\alpha\in[-0.42,-0.35]$}
\label{ris10}
\end{figure}
\par
\begin{figure}[H]
 \center{\includegraphics [width=0.8\linewidth]{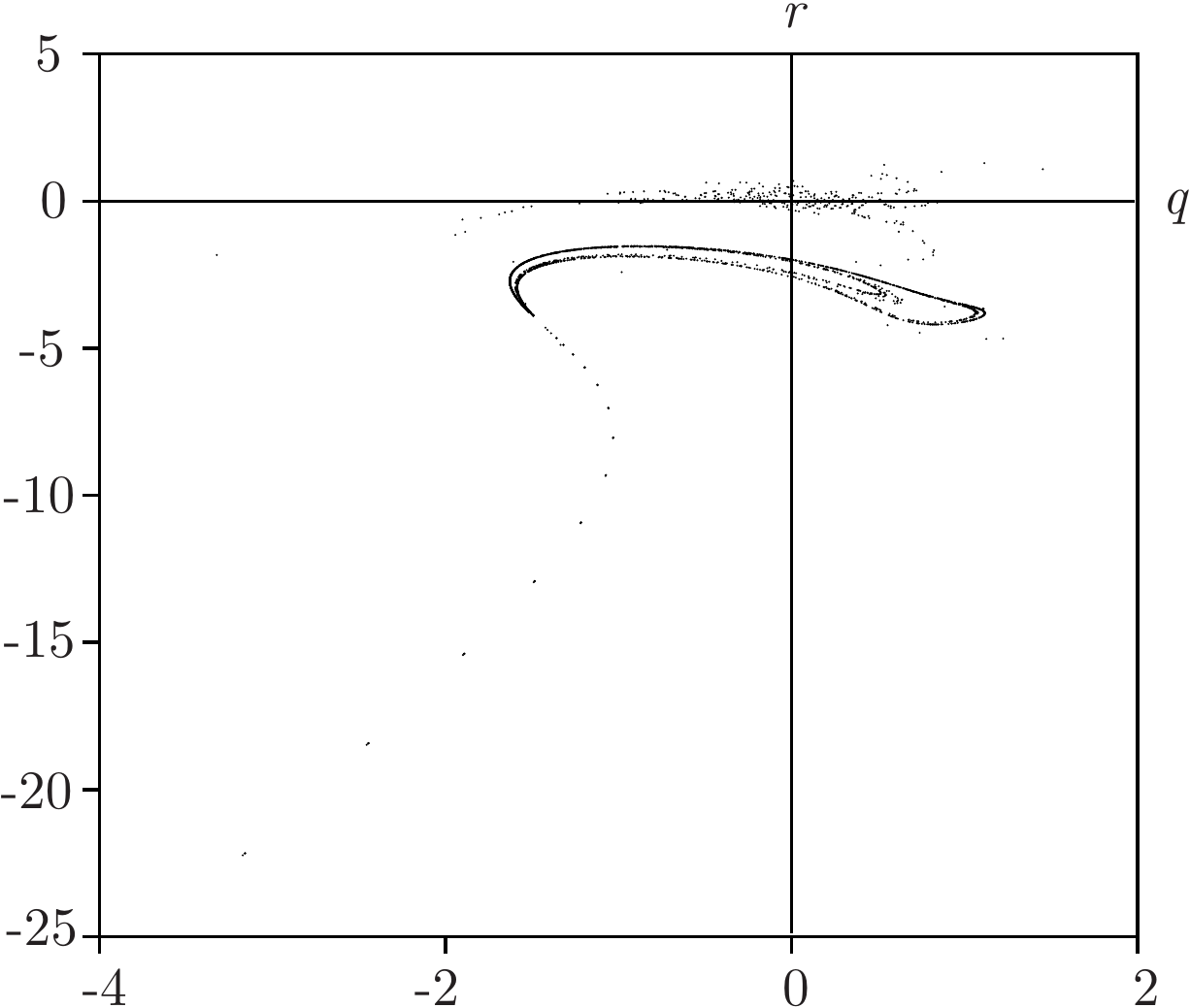}}\\
\begin{minipage}[h]{0.49\linewidth}
\center{\includegraphics [width=0.9\linewidth]{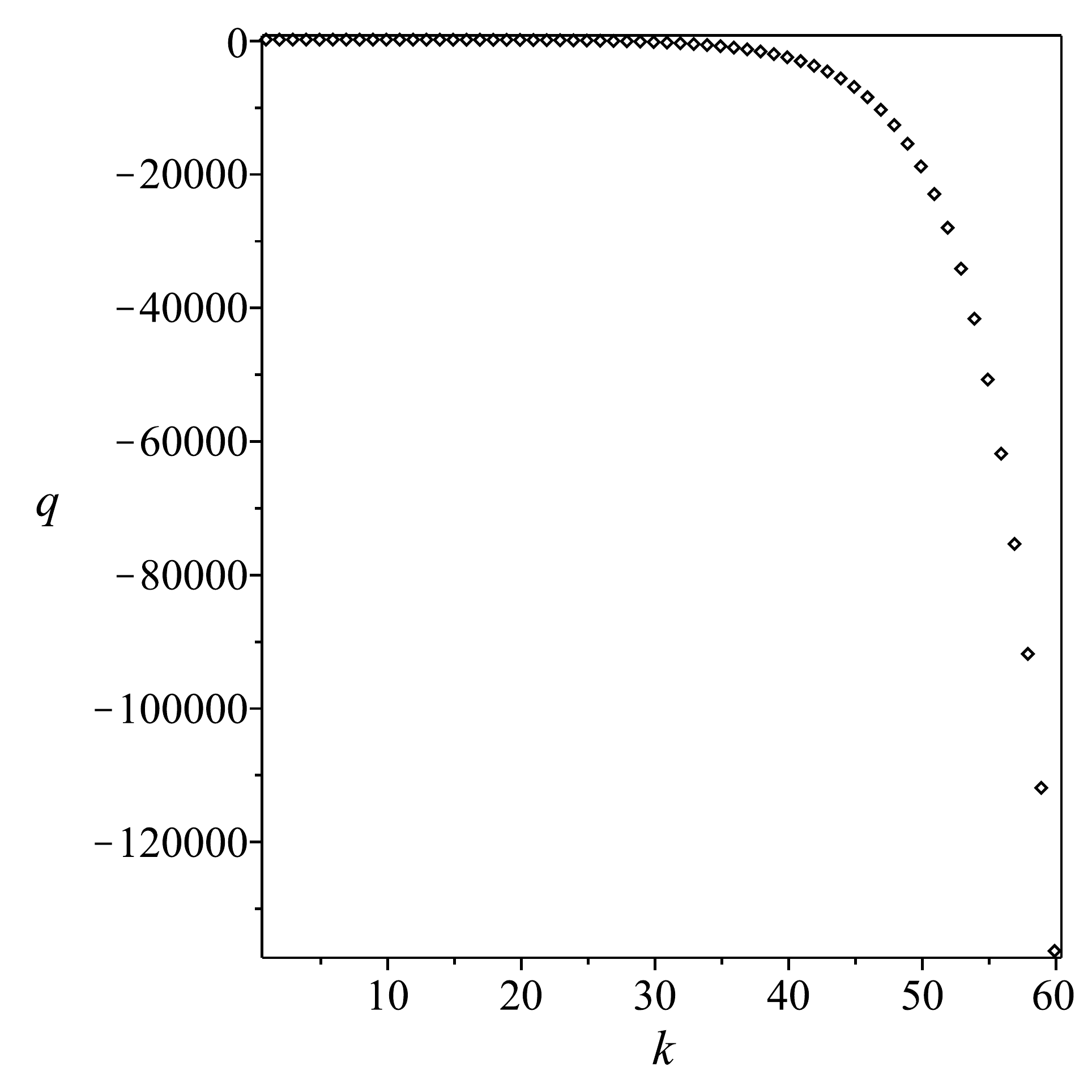}}
\end{minipage}
\hfill
\begin{minipage}[h]{0.49\linewidth}
\center{\includegraphics [width=0.9\linewidth]{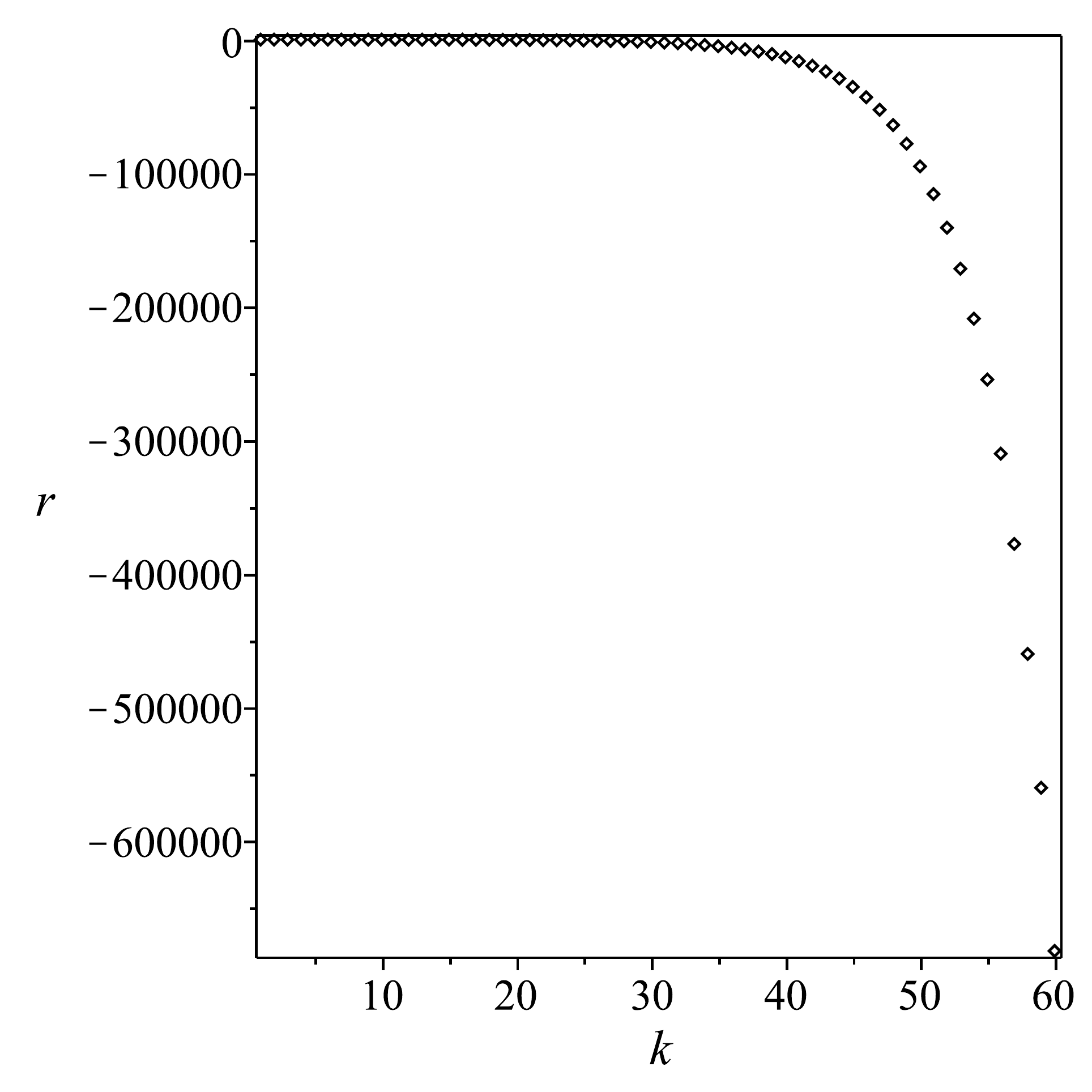}}
\end{minipage}
\caption{Poincar\'{e} section data and graphics of $q_k$ and $r_k$ at $\alpha=-0.45$}
\label{ris9}
\end{figure}

\noindent
As above, we can study the noncompact attractive curve given in Fig.\ref{ris9} at $\alpha=-0.45$.
\par\noindent
At $k=0\ldots 60$ the best fitting of these noncompact trajectories is given by six order polynomials
\[
q(t)=-0.0001\,t^6+0.0146\,t^5+\cdots\,,\qquad r(t)=-0.0005 t^6+0.073\,t^5+\cdots\,.
\]
This polynomial fitting is most likely not optimal in the four critical points case when a strange attractor is present in the Poincar\'{e} section.

\begin{figure}[H]
\center{\includegraphics [width=0.9\linewidth]{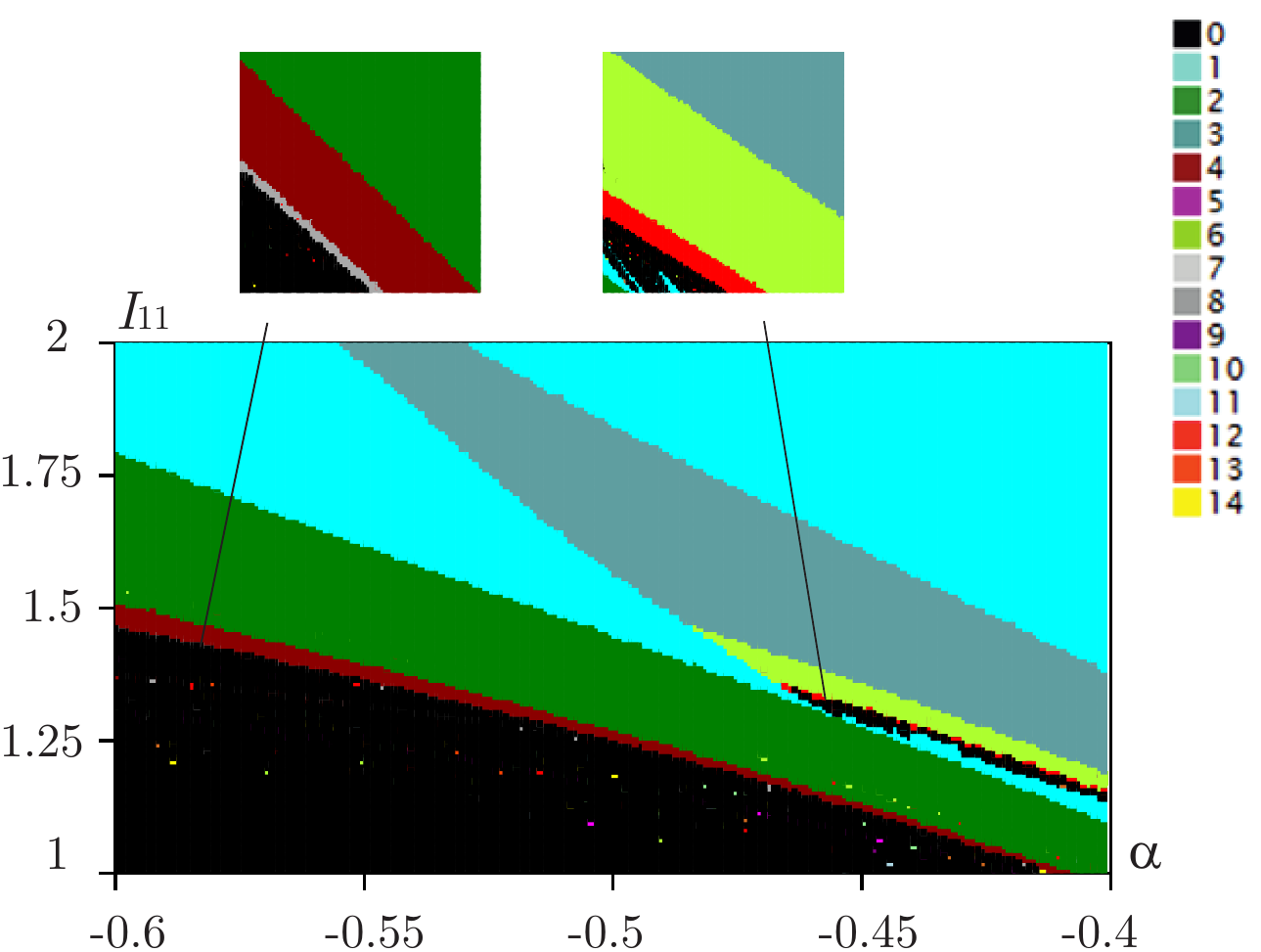}}
\caption{Chart of dynamical regimes on the $(I_{11}; \alpha)$ parameter plane}
\label{ris11}
\end{figure}

 In the case of two control parameters there also exists a useful visual representation of the system behavior through a chart of dynamical regimes on the parameter plane (bi-parametric sweep), where the domains of qualitatively distinct regimes are indicated by colors  \cite{kuz,sh10,st15}. An example of the chart of dynamical regimes on the $(\alpha, I_{11})$ parameter plane is given in Fig.\ref{ris11} at $\alpha\in [-0.6, -0.4]$, $I_{11}\in [1,2]$ and at fixed other entries of the tensor of inertia $I$ (\ref{i3}).
\par\noindent
The colored areas in Fig.\ref{ris11} correspond to stable cycles of the corresponding period (the color correspondence to the period is indicated in the upper right corner of the figure). Two black areas with colored splashes indicate the appearance of a strange attractor for the corresponding parameter values. Transition between these two chaotic regions occurs through cascades of period doubling 1-2-4-8 and 3-6-12, which indicates the Feigenbaum nature of the strange attractors.

\section{Conclusion}
The motion of the Euler top is mathematically described by geodesics of the left-invariant metric on rotation group $SO(3)$. In terms of physics, the Euler top is a rigid body moving about its center of mass without any forces acting on the body. It is one of the most studied systems in classical and quantum mechanics.

In \cite{bil} Bilimovich imposed a linear rheonomic constraint on the freely rotating rigid body, which preserved total mechanical energy, and as a result obtained an integrable time-dependent non-Hamiltonian system. The mechanical realization of this constraint proposed by Bilimovich is complex and most likely incorrect. Nevertheless, we suppose that classical effects such as free precession of the symmetric top, Feynman’s wobbling plate, tennis-racket instability, and the Dzhanibekov effect, attitude control of satellites by momentum wheels, or twisting somersault dynamics, have their counterparts in this abstract non-Hamiltonian system.

In this note, we discuss the nonlinear counterpart of Bilimovich's constraint, which does not preserve total mechanical energy and
which can be interpreted as rigid body control. The corresponding 3D non-Hamiltonian system with a time depending constraint is reduced to a periodically forced 2D non-autonomous system of differential equations, which allows us to apply all the well-known machinery to study the dynamic behavior of this system. We restrict ourselves by computing only a few Poincar\'{e} maps, portraits of quasi and strange attractors, Lyapunov exponents, Lyapunov dimension, and acceleration estimates. These computations highlight how the Poincar\'{e} map can be used to disambiguate and discover multiscale temporal dynamics, specifically the
coarse-grained dynamics resulting from fast-scale nonlinear control.

Of course, there are many other questions which have to be studied. For instance, applying the Poincar\'{e} maps method and time piecewise feedback control laws on Bilimovich's constraint for stabilization of the limit cycles, avoidance of areas with chaotic motion (strange attractors),  achievement of areas with acceleration for this periodically forced, non-autonomous, non-Hamiltonian system similar to \cite{bbm18,bk18} can be the potential avenues for future studies.

We have found a new very instructive example of a nonholonomic system similar to the Duffing oscillator, R\"{o}ssler system, RC circuit, driven Brusselator, etc. This model system is related to the freely rotating rigid body in Hamiltonian mechanics, nonholonomic mechanics, and control theory simultaneously.

We are very grateful to the referees for a thorough analysis of the manuscript, constructive suggestions, and proposed corrections, which certainly lead to a more profound discussion of the results.

This work of A.V. Borisov and A.V. Tsiganov was supported by the Russian Science Foundation (project no.~19-71-30012) and performed at the Steklov Mathematical Institute of the Russian Academy of Sciences. The authors declare that they have no conflicts of interest.


\begin{thebibliography}{10}

\bibitem{ar13}
Ariel G., Engquist B., Kim S., Lee Y., Tsai R.,
\newblock{\em A multiscale method for highly oscillatory dynamical systems using a Poincar\'{e} map type technique}
J. Scientific Comp., v.54 (2-3), pp.247-268, 2013.

\bibitem{bil} Bilimovich A. D.
\newblock{\em Sur les syst\`{e}mes conservatifs, non holonomes
avec des liaisons d\'{e}pendantes du temps}, Comptes Rendus Acad. Sci. Paris, v.156, pp.12-18, 1913.




\bibitem{bbm18}
 Bizyaev I A, Borisov A V, Mamaev I S.
 \newblock{\em Exotic Dynamics of Nonholonomic Roller Racer with Periodic Control},
 Regular and Chaotic Dynamics, v.23, pp. 983-994, 2018.



\bibitem{bl}
Bloch A.M.,
\newblock{ Nonholonomic mechanics and control. Interdisciplinary applied mathematics}, vol. 24. New York, NY: Springer-Verlag, 2003.



\bibitem{bk18}
 Borisov A.V., Kuznetsov S.P.,
 \newblock{\em Comparing Dynamics Initiated by an Attached Oscillating
Particle for the Nonholonomic Model of a Chaplygin Sleigh and for a Model with Strong Transverse and Weak Longitudinal Viscous Friction Applied at a Fixed Point on the Body},
 Regular and Chaotic Dynamics, v. 23, pp. 803-820, 2018.


\bibitem{bt20}
 Borisov A. V., Tsiganov A.V.,
 \newblock{\em On rheonomic nonholonomic deformations of the Euler equations proposed by Bilimovich},
 accepted to Theor. Appl. Mechanics, 2020.

 \bibitem{k19}
 Bramburger J.J., Kutz J. N.,
 \newblock{\em Poincar\'{e} Maps for Multiscale Physics Discovery and Nonlinear Floquet Theory},
 arXiv:1908.10958, 2019.


 \bibitem{k19a}
Brunton S. L., Kutz J. N.,
\newblock{ Data-driven science and engineering: Machine learning, dynamical
systems and control}, Cambridge University Press, Cambridge UK, 2019.

\bibitem{drag}
 Dragovi\'{c}V., B. Gaji\'{c} B.,
 \newblock{\em Hirota–Kimura type discretization of the classical nonholonomic Suslov problem}, Regul. Chaotic Dyn., v.13), pp. 250-256, 2008.

\bibitem{fed09}
 Fedorov Yu.N.,  Maciejewski A.J.,  Przybylska M.,
\newblock{\em The Poisson equations in the nonholonomic Suslov problem: integrability, meromorphic and hypergeometric solutions},
Nonlinearity, v. 22, pp. 2231-2259, 2009.

\bibitem{fer}
 Fernandez O.E., Bloch A.M., Zenkov D.V.,
 \newblock{\em The geometry and integrability of the Suslov problem}, J. Math. Phys.,2014, vol.55, no.11, 112704, 14pp.

\bibitem{nar14}
Garcia-Naranjo L.C., Maciejewski A.J., Marrero J.C., Przybylska M.,
\newblock{\em The inhomogeneous Suslov problem},
Phys. Lett. A, v.378, pp. 2389-2394, 2014.

\bibitem{grig}
Grigoryan A.T., Fradlin B.F.,
\newblock{
Scientific heritage of G.K. Suslov's school on analytical mechanics and its development in the research of Yugoslav scientists}, Mathematical Institute, Beograd, 1977.

\bibitem{grab09}
Grabowski J., de Le\'{o}n M., Marrero J.C., de Diego D.M.,
\newblock{\em Nonholonomic constraints: a new viewpoint},
J. Math. Phys., v.50, 013520, 2009.


\bibitem{hen}
H\'{e}non M.,
\newblock{\em On the numerical computation of Poincar\'{e} maps}, Physica D, v. 5, pp. 412–414, 1982.

\bibitem{kar}
Karapetian A.V.,
\newblock{\em On realizing nonholonomic constraints by viscous friction forces and Celtic
stones stability}, J. Appl. Math. Mech., v. 45, pp. 42-51, 1981.

\bibitem{kol04}
Rios P.M., Koiller J.,
 \newblock{Non-holonomic systems with symmetry allowing a conformally symplectic reduction},
 New Advances in Celestial Mechanics and Hamiltonian Systems, Springer US, 2004.

\bibitem{koz}
Kozlov V.V,
\newblock{\em The problem of realizing constraints in dynamics},
J. Appl. Math. Mech., v. 56, pp. 594-600, 1992.

 \bibitem{kuz}
 Kuznetsov, Yu. A., Meijer H. G. E.,
 \newblock{Numerical Bifurcation Analysis of Maps
From Theory to Software}, Cambridge University Press, 2019.

\bibitem{dl}
de Le\'{o}n M.,
\newblock{\em A historical review on nonholonomic mechanics}, Rev. R. Acad. Cienc. Exactas
F\'{\i}s. Nat. Ser. A Math. RACSAM, v. 106, no. 1, 191-224, 2012.

\bibitem{nf72}
 Neimark J., Fufaev N., Dynamics of Nonholonomic Systems, Transactions of Mathematical Monographs,
v. 33, AMS, Providence, RJ, 1972.

\bibitem{mac}
Maciejewski A.J., Przybylska M.,
\newblock{\em Nonintegrability of the Suslov problem}, J. Math. Phys., v.45, pp.1065–1078, 2004.

\bibitem{mah12}
Mahdi A., Valls C.,
\newblock{\em Analytic non-integrability of the Suslov problem}, J. Math. Phys.,2012, v.53, 122901, 2012.

\bibitem{par89}
Parker T.S., Chua L.O.,
\newblock{Practical numerical algorithms for chaotic systems}, Springer-Verlag, New York, 1989.

\bibitem{poi}
Poincar\'{e} H.,
\newblock{ Les methodes nouvelles de la m\'{e}echanique c\'{e}eleste},
Gauthier-Villars, Paris, 1899.

\bibitem{put}
 Putkaradze V., Rogers S.,
 \newblock{\em Constraint control of nonholonomic mechanical system},
 J. Nonlinear Science, v. 28, pp. 193-234, 2018.

\bibitem{sh10}
Shilnikov L. P.,
\newblock{ Method of Qualitative Theory in Nonlinear Dynmics},
Part 1-2, Higher Education Press, Beijing, China, 2010.

\bibitem{st15}
Strogatz S.,
\newblock{ Nonlinear dynamics and chaos: with applications to physics, biology, chemistry, and engineering}, Westview Press, Boulder CO, 2015.

\bibitem{sus}
G. K. Suslov, Fundamentals of analytical mechanics, Vol. 2, 1902, Kiev.


\bibitem{tuk}
Tucker W.,
\newblock{\em Computing accurate Poincar\'{e} maps}, Physica. D: Nonlinear Phenomena,
v.171, pp. 127-137, 2002.

\bibitem{vag}
Vagner V.V.,
 \newblock{\em A geometric interpretation of nonholonomic dynamical systems}, Tr. Semin. Vectorn. Tenzorn. Anal.,v.5, pp.301–327, 1941.


\end{thebibliography}
\end{document}